\documentclass[aps,prl,reprint,longbibliography]{revtex4-2}
\usepackage{amssymb}
\usepackage{amsmath}
\usepackage{color}
\usepackage{epsfig}
\usepackage{subfigure}
\usepackage{hyperref}
\usepackage{verbatim}
\usepackage{dcolumn}
\usepackage{natbib}

\newcommand{\corr}[1]{{\color{black} #1}}

\begin{document}

\title{Information-to-work conversion in single molecule experiments: \\ from discrete to continuous feedback}
\author{Regina K. Schmitt}
\affiliation{Department of Physics and NanoLund, Lund University, Box 188, SE-221 00 Lund, Sweden}
\author{Patrick P. Potts}
\affiliation{Department of Physics and NanoLund, Lund University, Box 188, SE-221 00 Lund, Sweden}
\author{Heiner Linke}
\affiliation{Department of Physics and NanoLund, Lund University, Box 188, SE-221 00 Lund, Sweden}
\author{Marc Rico-Pasto}
\affiliation{Small Biosystems Lab, Condensed Matter Physics Department, Universitat de Barcelona, C/Marti i Franques 1, 08028 Barcelona, Spain}

\author{Jonas Johansson}
\affiliation{Department of Physics and NanoLund, Lund University, Box 188, SE-221 00 Lund, Sweden}
\author{Felix Ritort}
\affiliation{Small Biosystems Lab, Condensed Matter Physics Department, Universitat de Barcelona, C/Marti i Franques 1, 08028 Barcelona, Spain}
\author{Peter Samuelsson}
\affiliation{Department of Physics and NanoLund, Lund University, Box 188, SE-221 00 Lund, Sweden}

\begin{abstract}
We theoretically investigate the extractable work in single molecule unfolding-folding experiments with applied feedback. Using a simple two-state model, we obtain a description of the full work distribution, from discrete to continuous feedback. The effect of the feedback is captured by a detailed fluctuation theorem, accounting for the information aquired. We find analytical expressions for the average work extraction as well as an experimentally measurable bound thereof, which becomes tight in the continuous feedback limit. We further determine the parameters for maximal power, or rate of work extraction. While our two-state model only depends on a single, effective transition rate, we find quantitative agreement with Monte Carlo simulations of DNA hairpin unfolding-folding dynamics. 
\end{abstract}

\maketitle

\emph{Introduction---}
The ability to manipulate and measure systems at the nanometer and the piconewton scale has driven the need to understand systems that are subject to large
fluctuations, out of thermal equilibrium.  Stochastic thermodynamics provides the theoretical framework for describing such systems. A cornerstone is provided by fluctuation theorems (FTs) \cite{Harris2007,Esposito2009rmp,Jarzynski2011,Seifert2012c,Mansour2017,campisi:2011}, most prominently the Crooks FT 
\cite{Crooks1998,Crooks1999,Crooks2000} and the Jarzynski equality \cite{Jarzynski1997a,Jarzynski1997}, which leads to the second law, $\left\langle
W\right\rangle\geq\Delta F$. Hence, in work generating processes, with $\Delta F\leq 0$ the work extracted along a single trajectory, $-W$, can be larger than the free energy difference
$|\Delta F|$. Taking advantage of such transient violations (TVs) of the second law, information and feedback (FB) may be used to increase the average extractable work \cite{Sagawa2008,CaoFeito2009,Sagawa2010,Horowitz2010,Ponmurugan2010,Sagawa2011a,Sagawa2012,Lahiri2012,Abreu2012,Ashida2014,Horowitz2014,Horowitz2014prx,Wachtler2016,potts:2018}. For a single measurement with subsequent FB, Sagawa and Ueda \cite{Sagawa2008} found a generalization of the second law, $\left\langle W\right\rangle\geq \Delta F-k_\text BT\langle I \rangle$, with the thermal energy $k_\text BT$ and the mutual information between system state and measurement outcome $\langle I \rangle$. This inequality was experimentally verified using an
optically trapped colloidal particle \cite{Toyabe2010}. Similar inequalities where found for consecutive discrete measurements
\cite{Ponmurugan2010,Horowitz2010,Lahiri2012,Sagawa2012,Fujitani2010,Ashida2014} but its extension to the continuous FB limit proved to be problematic as  $\langle I \rangle$ tends to diverge \cite{Sagawa2012,Horowitz2014}. A remedy to this problem was provided in Ref.~\cite{potts:2018}, where a recipe for deriving fluctuation theorems in the presence of measurement and feedback was given, highlighting the fact that many such theorems exist.\\
\begin{figure}[h]
  \includegraphics[width=1\columnwidth]{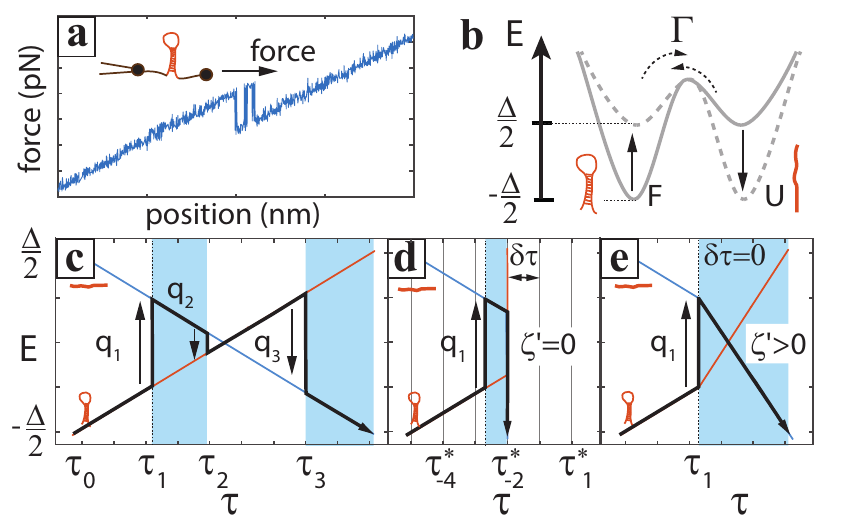} 
	\caption{(a) Force-position trace for single DNA hairpin experiment, with force jumps showing folding/unfolding events (see text). Inset: DNA molecule held between a micro pipette and movable optical tweezers, with a force applied. (b) Schematics of the molecule free energy landscape for initial (solid) and final (dashed) times of the protocol. The two molecular states $\text U$ (unfolded) and $\text F$ (folded) shown at their corresponding local energy minima.  (c)-(e) Energy-time trajectories in state space with transitions at times $\tau_n$, denoted by vertical arrows. The dissipated (dimensionless) heat at the transitions is $q_n=\pm 2\tau_n$. Regions in time where work is extracted are shaded blue. In (d) measurements and FB are performed at times $\tau_n^*$, in intervals $\delta\tau$. At the measurement detecting the first transition $\text F \rightarrow \text U$ the states are shifted, with infinite drive speed, $\zeta'=0$, to their final energies. In (e) measurement and FB is continuous ($\delta\tau=0$), with non-zero $\zeta'$: Directly after a first transition, the drive speed is increased from $\zeta$ to $\zeta'<\zeta$. Red and blue lines in (c-e) stand for the folded and unfolded energy branches.}
\label{figsys}
\end{figure}
Experimentally, central results of stochastic thermodynamics were verified in a number of architectures. Examples are the verification of Landauer's principle
\cite{Landauer1961} using optical tweezers \cite{Berut2012} and a virtual potential feedback trap \cite{Jun2014}, implementations of Maxwell's Demon \cite{Maxwell1871}
and Szilard's engine \cite{Szilard1929} using a colloidal particle \cite{Toyabe2010}, single-electron boxes \cite{Koski2014,Koski2015,Chida2017},
superconducting circuits \cite{Cottet2017,Masuyama2018,Naghiloo2018}, as well as thermal light \cite{Vidrighin2016}, and the verification of FTs and the
determination of free energies using optically trapped particles \cite{Wang2002a,Trepagnier2004,Carberry2004,Alemany2012,Hoang2018a} and quantum dots
\cite{Hofmann2016,Hofmann2017}. Of particular interest are experiments based on single molecule force spectroscopy
(SMFS) \cite{Liphardt2001,Ritort2002,Liphardt2002,Collin2005,Manosas2005,ManosasMossa2009a,ManosasMossa2009b,Dieterich2015,Dieterich2016}, providing unique possibilities of simultaneous force and molecular extension measurements in a biological system, making work directly accessible, see Fig. \ref{figsys}. SMFS on DNA/RNA
hairpins was used to verify the Jarzynski relation \cite{Liphardt2002} and the Crooks FT \cite{Collin2005}, as well as to investigate a continuous-time version of Maxwell's demon \cite{ribezzi:2019nat,ribezzi:2019}. In a recent work by some of us, the effect of feedback on \corr{dissipation reduction and} improved free energy determination was investigated in single molecule pulling experiments \cite{rico:2020}.
    
In this letter, we theoretically investigate the extraction of work in a SMFS experiment on DNA hairpins, providing a detailed understanding of information-to-work conversion for FB ranging from discrete to continuous. We consider a DNA strand that is attached at both ends, see Fig.~\ref{figsys}\,(a). Its ends are then pulled apart with a constant velocity. During this process, measurements of the system state are performed. As soon as the DNA strand is found to be unfolded, the velocity is increased, see Fig.~\ref{figsys}\,(d) and (e), resulting in the extraction of work.
We model the experiment with a single parameter, two-state system \cite{Ritort2002,Ritort2004,Chvosta2007,Subrt2007} coupled to a single heat bath, comparing well to detailed Monte-Carlo simulations \cite{Manosas2007b} of the full system. We show that going from discrete to continuous FB, the amount of extractable work increases, in agreement with Ref.~\cite{rico:2020}. Based on a detailed FT which circumvents problems encountered in continuous and error-free measurements \cite{potts:2018}, we derive integral FTs and a bound for the extractable work, becoming tight in the limit of continuous FB.
We moreover identify optimal parameters for work extraction and power production.

\emph{Two-state model---}
Dynamical SMFS of DNA hairpin experiments are well described by Monte-Carlo simulations with detailed DNA models \cite{AlemanyThesis}.  However, the key dynamical features of the hairpin experiments can be captured by simple two-state models \corr{\cite{Manosas2005}}. Such two-state models often allow for analytical treatments of the full work distribution \cite{Ritort2002,Ritort2004,Chvosta2007,Subrt2007}, providing compelling and transparent pictures of the underlying physics. Here we  focus on the simplest possible two-state model that captures the full dynamics with an effective transfer rate. Key results are compared to a detailed DNA Monte Carlo simulation, discussed below.

We first consider the system in absence of FB. The two system states, with the molecule folded (F) or unfolded (U) see Fig. \ref{figsys} (b), have energies driven linearly in time as $E_\text F(t)=-E_\text U(t)=\kappa t$ where $\kappa$ is the constant energy velocity and the energies are degenerate (and set equal to zero) at $t=0$. The drive protocol is symmetric, such that $|E_\text F-E_\text U|=\Delta$ both at the beginning $(t=-\Delta/2\kappa)$ as well as at the end of the protocol $(t=\Delta/2\kappa)$, i.e., there is no free energy difference between the initial and final state, $\Delta F=0$. Throughout the paper, we keep $\Delta$ fixed which implies that protocols with different velocities $\kappa$ take a different amount of time. We consider the experimentally relevant \cite{Collin2005} limit $k_\text BT\ll\Delta$, where the system is initially in state $\text F$ (in thermal equilibrium) and ends in state $\text U$. The transitions between $\text F$ and $\text U$ are thermally activated, with time-dependent rates $\Gamma e^{\pm t\kappa /(k_\text B T)}$, Fig.~\ref{figsys}\,(b), which fulfill local-in-time detailed balance by construction. The constant attempt rate $\Gamma$ depends on system parameters, e.g., the height of the energy barrier separating F and U. We note that in DNA-pulling experiments, the condition $\Delta F=0$ is usually not fulfilled. However, a finite $\Delta F$ can be accounted for by a constant shift of the extracted work,  $W \rightarrow W + \Delta F$.  Moreover, the symmetric kinetic rates correspond to a barrier located half distance between F and U \cite{rico:2020}.

The dynamics of the state occupation probabilities $P_\text F(t)$, $P_\text U(t)=1-P_\text F(t)$ is described by a rate equation with time-dependent rates. Introducing the dimensionless time $\tau=\kappa t/(k_\text B T)$, and the dimensionless, effective attempt rate $\zeta=k_{\rm B}T\Gamma/\kappa$, we have
\begin{equation}
\frac{dP_\text F(\tau)}{d\tau}+2\zeta\cosh(\tau) P_\text F(\tau)=\zeta e^{-\tau}, 
\label{rateEQ}
\end{equation}
showing that the dynamics is completely governed by $\zeta$. The solution to Eq.~(\ref{rateEQ}) for $\tau>\tau_0$, with $\tau_0=-\Delta/(2k_\text B T)$ the initial time and $P_\text F(\tau_0)=1$, can be written as
\begin{eqnarray}
P_\text F(\tau)=1-\zeta \int_{\tau_0}^{\tau} ds ~e^{s+2\zeta [\sinh s- \sinh \tau]},
\label{P1exp}
\end{eqnarray} 
We note that for $\zeta\gg 1$ we recover the quasi-static limit with multiple transitions $\text F\leftrightarrow \text U$, giving the equilibrium result $P_\text F(\tau|\zeta \gg 1)=1/(1+e^{2\tau})$. For $\zeta\ll 1$ we enter the rapid drive regime where only a single transition $\text F\rightarrow \text U$ takes place and $P_\text F(\tau|\zeta\ll 1)=\exp[-\zeta\exp(\tau)]$  \cite{SI}. 

\begin{figure}
\includegraphics[width=0.5\textwidth]{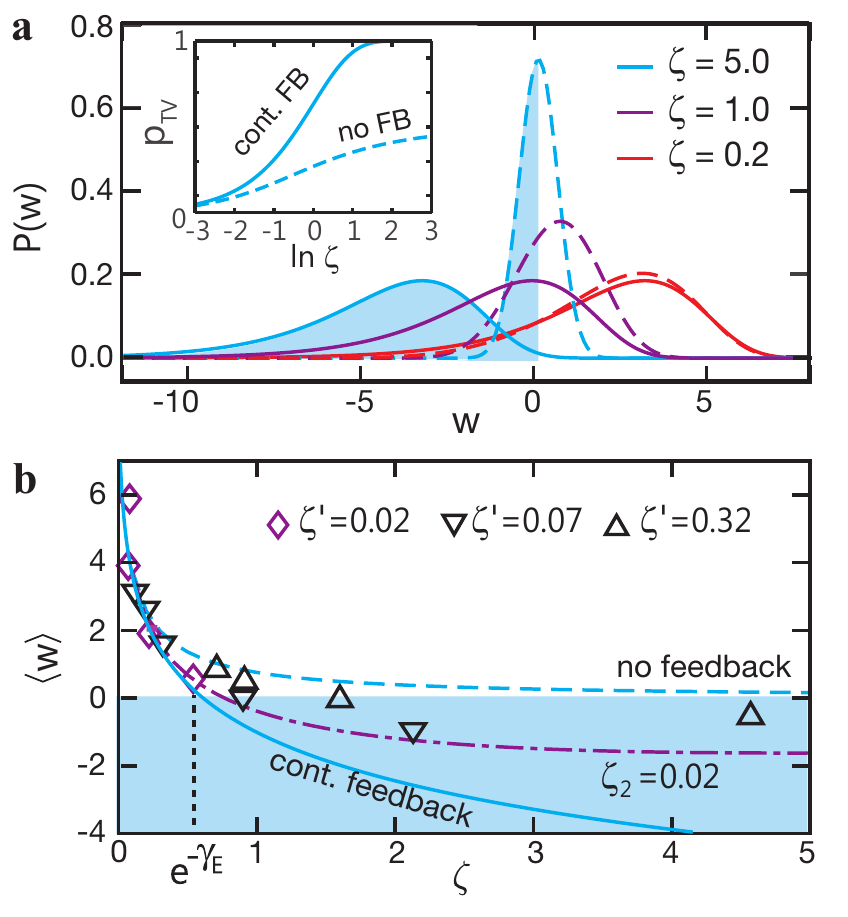}
\caption{(a) Probability distribution of work $w$ for protocols with no FB, (dashed lines) and continuous FB (solid lines) for three different effective rates $\zeta$. Continuous FB shifts the distribution towards negative $w$, most notably in the quasi-static regime $\zeta\gg 1$. The inset shows the fraction $p_{\text{TV}}$ of TV contributions without feedback (dashed line) and continuous feedback (solid line), as a function of $\zeta$. (b) Average work from dynamical simulations of continuous FB with a finite rate $\zeta'$ after the first transition (see text), for different sets of  $\zeta,\zeta'$  (empty symbols). The work without FB, Eq. (\ref{w}), (dashed line) and continuous FB for $\zeta' \rightarrow 0$, Eq. (\ref{w_cM}), (solid line) and $\zeta'=0.02$ (dash-dotted line) for the two-state system are shown for reference.}
\label{distplots}
\end{figure}

{\em Work distribution ---} Because the internal energy of the molecule is the same at the beginning and at the end of each trajectory, the first law of thermodynamics (which holds on each trajectory) results in $W=-Q$, where $W$ is the work performed on the system and $Q$ the heat absorbed from the environment. In the following, we will work with the dimensionless quantities $w=W/(k_\text B T)$ and $q=Q/(k_\text B T)$. A given trajectory with $N$ state transitions is completely determined by the set of transition times $\{ \tau_n\}_{n=1}^N$.  Moreover, a transition at $\tau_n$, with $n=1,3,...,N ~  (n=2,4,...N-1)$ for $\text F \rightarrow \text U~(\text U \rightarrow \text F)$, gives rise to a \corr{transferred heat $q_n=-2\tau_n~(q_n=2\tau_n)$}, equal to the energy difference between the two states, see Fig.~\ref{figsys}\,(c) (note that the system always starts in the folded state). The total work along the trajectory is then $w=2(\sum_{n=1,3,..}^{N}-\sum_{n=2,4,...}^{N-1})\tau_n$ and the distribution of the work performed, $P(w)$, is thus directly obtained from the distribution of transition times; the derivation for arbitrary $\zeta$ is presented in the supplementary information \cite{SI}. In the quasi-static limit the distribution becomes a shifted Gaussian 
\begin{eqnarray}
P_{\text{nf}}(w|\zeta\gg 1)=\frac{\sqrt{\zeta}}{\pi}\exp \left[-\frac{\zeta}{\pi}\left(w-\frac{\pi}{4\zeta}\right)^2\right],
\label{pwqs}
\end{eqnarray}
while in the rapid drive regime we find
\begin{eqnarray}
P_{\text{nf}}(w|\zeta\ll 1)=\frac{1}{4K_1(2\zeta)}\exp\left[\frac{w}{2}-2\zeta\cosh\left(\frac{w}{2} \right) \right],
\label{pwrap}
\end{eqnarray}
where $K_1(\zeta)$ is a modified Bessel function of the second kind and the subscript {nf} denotes \textit{no feedback}. We stress that $P_\text {nf}(w)$ for any $\zeta$ obeys Crooks fluctuation theorem \cite{Crooks1998,Crooks1999}, which in our symmetric case reads $P_\text{nf}(w|\zeta)/P_\text{nf}(-w|\zeta)=e^{w}$. 

As is clear from Fig.~\ref{distplots}\,(a), decreasing $\zeta$ shifts $P_\text{nf}(w)$ towards more positive $w$. In particular, the average work 
\begin{eqnarray}
\langle w \rangle_\text{nf}=\zeta \frac{\pi^2}{4}\left[J_{0}(2\zeta)J_1(2\zeta)+Y_0(2\zeta)Y_1(2\zeta)\right],
\label{w}
\end{eqnarray}
is always positive, see Fig~\ref{distplots}\,(b). Here $J_{\nu}(x)$ [$Y_{\nu}(x)$], with $\nu=0,1$, is a Bessel function of the first [second] kind and $\langle ... \rangle_\text{nf}=\int dw ... P_\text{nf}(w)$. However, for any $\zeta$ there is a non-zero probability for transient violations (TV) of the second law; the fraction of TV-trajectories, $p_{\text TV}$, goes from $0.5$ in the quasi-static limit towards zero in the rapid regime, see the inset in Fig.~\ref{distplots}\,(a).

{\em FB-enabled work extraction ---} In order to extract work on average, we consider the use of FB to increase the fraction of TV-trajectories.
To this end, we consider an ideal FB protocol with repeated, error-free, non-invasive measurements of the system state. These measurements are performed at times $\tau_m^*=\text{sgn}(m) \delta\tau(|m|-1/2)$ for integers $m=\pm 1, \pm2,..$ (for $|\tau_m^*|<|\tau_0|$), i.e., they are separated in time by $\delta\tau$ and are situated symmetrically around $\tau=0$. Since the measurements are performed at discrete times, we call this protocol a \textit{discrete} FB protocol \cite{discrete}. Initially, at $\tau=\tau_0$, the system is in state $\text F$. The energy levels are then moved with velocity $\kappa$ (effective attempt rate $\zeta$). For every measurement, the possible outcomes are $\text F$ and $\text U$. If the system is found in $\text U$, the system is instantaneously taken to its end position $E_\text F-E_\text U=\Delta$ (i.e., the velocity is taken to infinity,  $\zeta\rightarrow 0$) and the protocol is ended without further state transitions. If the system instead is found in $\text F$ no FB is performed and the system evolves, according to Eq. (\ref{rateEQ}), to the next measurement.

\begin{figure}
	\centering
\includegraphics[width=1\columnwidth]{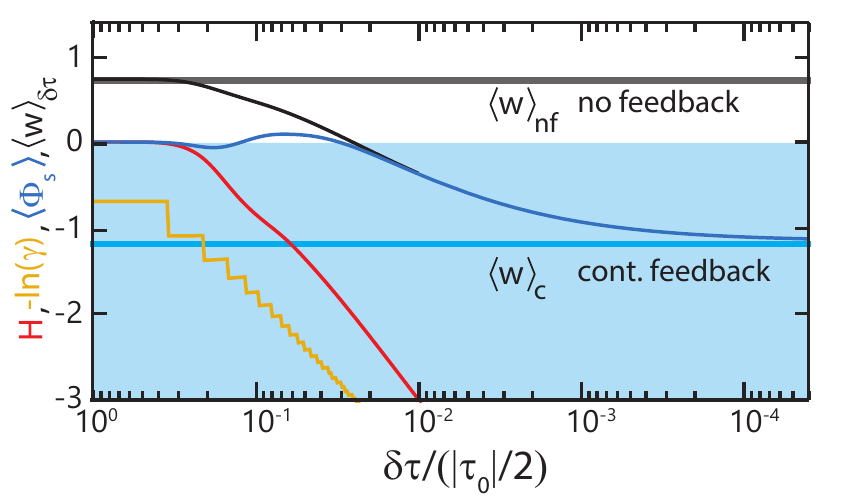}
	\caption{Average work $\langle w \rangle_{\delta\tau}$  for the discrete FB protocol as a function of the time between the measurements, $\delta\tau$, for $\tau_0=-10$. Note that the time axis goes from large to small $\delta \tau$. \corr{The work (black line) decreases with decreasing $\delta\tau$ before it becomes negative and eventually saturates at the continuous measurement result $\langle w \rangle_\text c$ (cyan line). Three different entropies (see text) $ H $ (Shannon, red), $\ln(\gamma)$ (logarithmic efficacy, yellow) and $\langle \Phi_\text s \rangle$ (blue)} are plotted, showing that while  $H$ and  $-\ln(\gamma)$ diverge in the continuous FB limit, $\langle \Phi_\text s \rangle$ is finite, constituting a tight bound on the extractable work.}
\label{manymeas}
\end{figure}

The resulting average work \cite{SI}, denoted  $\langle w \rangle_{\delta\tau}$, is plotted in Fig.~\ref{manymeas} as a function of $\delta\tau$, for a given $\zeta$. It is clear from the plot that $\langle w \rangle_{\delta\tau}$ decreases monotonically as $\delta\tau$ is reduced. A careful analysis shows that this holds for any $\zeta$ (not shown). In particular, the average work becomes negative, showing that for sufficiently small $\delta \tau$, work is extracted using FB. 

Interestingly, in the limit of ${\delta\tau} \rightarrow 0$, the average work saturates at a constant value. In this limit the FB protocol corresponds to a continuous monitoring of the system state, with a change to infinite drive speed immediately when the first transition $\text F \rightarrow \text U$ occurs. From the known distribution of $\tau_1$, the first unfolding time \cite{SI}, and recalling that the heat $q_1$ absorbed at the transition is equal to $-2\tau_1$, we can directly write down the distribution of performed work as
\begin{eqnarray}
P_\text{c}(w)=\frac{\zeta}{2} e^{w/2-\zeta e^{w/2}},
\label{pW_cM}
\end{eqnarray}
a Gumbel distribution (see Fig.~\ref{distplots}). Here the subscript \textit{c} stands for \textit{continuous monitoring} and corresponds to $\delta\tau\rightarrow 0$. The average work (Fig~\ref{distplots}\,(b), continuous blue line), reads
\begin{eqnarray}
\left\langle w\right\rangle_\text{c}=-2(\ln{\zeta} +\gamma_\text {E}),
\label{w_cM}
\end{eqnarray}
with $\gamma_\text E \approx 0.577$ the Euler constant.  The average work decreases with increasing $\zeta$, becoming zero for $\zeta=e^{-\gamma_\text E}\approx 0.561$. For larger $\zeta$ we can thus achieve a net heat extraction from the bath. In fact, as is clear from Eq. (\ref{pW_cM}) and shown in Fig. \ref{distplots}, increasing $\zeta$ only shifts the entire $P_\text{c}(w)$ to smaller work values, without changing the shape of the distribution. As a result, the fraction of TV-trajectories increases towards unity with increasing $\zeta$, shown in the inset of Fig. \ref{distplots} (a). Note that for $\zeta \rightarrow 0$, i.e., for infinitely fast drive, no FB is performed and the expressions in Eqs. (\ref{pW_cM}) and (\ref{pwrap}) coincide.

From Eq. (\ref{w_cM}) we see that $\left\langle w\right\rangle_\text{c}$ diverges when $\zeta \rightarrow \infty$, in the quasi static regime. In reality, the work is bounded by $\Delta/k_\text B T\gg 1$.  An informative figure of merit is the work extraction per unit time, or power.  Performing the protocol takes the time $t_\text p(w)$  up to the first observed transition, given by
\begin{equation}
\label{eq:protocoltime}
 t_\text p=\frac{k_\text B T}{\kappa}(\tau_1-\tau_0)=\frac{\zeta}{2\Gamma}\left(w+\frac{\Delta}{k_\text B T}\right)\simeq\frac{\zeta}{2\Gamma}\frac{\Delta}{k_\text BT},
\end{equation}
where we used that $\Delta/(k_\text B T)\gg w$ in all cases of interest. The average power produced by the system then reads
\begin{equation}
\left\langle \frac{w}{t_\text p(w)}\right\rangle_\text c\simeq-\frac{k_\text BT}{\Delta}\frac{4\Gamma}{\zeta}\left(\ln{\zeta} +\gamma_\text{E}\right)>0,
\label{Maxpow_cMF}
\end{equation}
which is finite and maximal for $\zeta=e^{1-\gamma_\text E}\approx 0.65$. 

{\em Information bound on work extraction ---} To clarify the role of information in the FB-process, we consider a detailed fluctuation theorem  (FT) \cite{potts:2018} applicable to repeated, discrete FB with arbitrary $\delta \tau$, i.e. including continuous FB. The FT is formulated in terms of conditional probability distributions for work performed in a 'forward' and a 'backward' experiment.  The forward experiment, described above, is characterized by the protocol $\lambda_\text s$, where the drive speed is switched from $\kappa=\Gamma/\zeta$ to infinity at $\tau_\text s^*$, upon measuring for the first time the system in state U. Hence,  $\tau_\text s^*, \zeta$ and $\delta \tau$ completely determine $\lambda_\text s$. The joint probability for a given value of work $w$ and a switching time $\tau_\text s^*$ is denoted $P(w,s)$. In the backward experiment, the  time-reversed protocol ${\lambda}^{\dagger}_\text s$ is applied with probability $p_\text s=\int dwP(w,s)$. This protocol initiates the system in state U (in thermal equilibrium), at $E_\text U=-E_\text F=-\Delta/2$, immediately takes the system to energy $E_\text U=-E_\text F=-k_BT\tau_\text s^*$ and then shifts the energies with speed $\kappa$ in the opposite direction compared to the forward experiment. During the finite drive speed, measurements are performed with the same interval $\delta \tau$ as in the forward experiments. \corr{The first measurement is performed when changing speed and necessarily results in U. Considering only trajectories where all subsequent measurements result in F} \cite{SI}, we have the FT 
\begin{equation}
P(w|s)={P}^\dagger(-w|s)e^{w-\Phi_\text s}, \quad \Phi_\text s=\ln(p_\text s/{p}^\dagger_\text s).
\label{FT}
\end{equation}
We note that a similar FT was employed in Ref.~\cite{rico:2020}, cf. Eq. (4) therein.
Here the forward conditional probability for work $P(w|s)=P(w,s)/p_\text s$ and ${P}^\dagger(w|s)$ is the corresponding backward conditional probability given that the protocol ${\lambda}^{\dagger}_\text s$ is applied and all measurements result in F. The fraction of backward trajectories under ${\lambda}^{\dagger}_\text s$ that give rise to measurement outcomes F \corr{for all but the first measurement} is denoted ${p}^\dagger_\text s$. Note that while $\sum_sp_\text s=1$ by construction, the quantity $\sum_s {p}^\dagger_\text s\equiv \gamma$, the efficacy of the protocol \cite{Sagawa2010,Sagawa2012}, is typically not unity.

From Eq. (\ref{FT}) we get the integral fluctuation theorems $\langle e^{-w} \rangle_{\delta \tau}=\gamma$ and $\langle e^{-w+\Phi_\text{s}} \rangle_{\delta\tau}=1$ where from the latter theorem, via Jensen's inequality, we get the modified second law
\begin{equation}
\langle w \rangle_{\delta \tau} \geq  \langle \Phi_\text s \rangle_{\delta \tau} =\sum_s p_\text s \ln(p_\text s/p_\text s^{\dagger}),
\label{2law}
\end{equation}
providing a bound on the extractable, average work. Two important remarks can be made about Eq. (\ref{2law}). First, the entropy, or information, term $\langle \Phi_\text s \rangle_{\delta \tau}$ depends only on probabilities for measurement outcomes, allowing one to experimentally determine the work bound. Second, $\langle \Phi_\text s \rangle_{\delta \tau}$ is finite in the continuous FB limit, ${\delta \tau}\rightarrow 0$, in contrast to e.g., the (negative) Shannon entropy $H=\sum_s p_\text s \ln(p_\text s)$ and efficacy $\gamma$, as also illustrated in Fig. \ref{manymeas}. In fact, we find that  $\langle \Phi_\text s \rangle_{\text c}=\langle w \rangle_{\text c}$, i.e., in the continuous FB limit the bound on the extractable work in Eq. (\ref{2law}) is tight. As discussed in Ref.~\cite{potts:2018}, this is because the measurement outcomes contain the full knowledge of the entropy production. 
 
\paragraph{Comparison to dynamical DNA simulations --- }
To emphasize the relevance of our two-state model to unfolding-folding experiments with DNA hairpins, we extend our idealized, continuous FB model to account for finite driving speed after the first unfolding event. That is, we consider a protocol, $\lambda_\text s(\zeta,\zeta')$ with effective transfer rates $\zeta$ and $\zeta'$ before and after the transition time respectively. 
\begin{figure}
    \includegraphics[width=1\columnwidth]{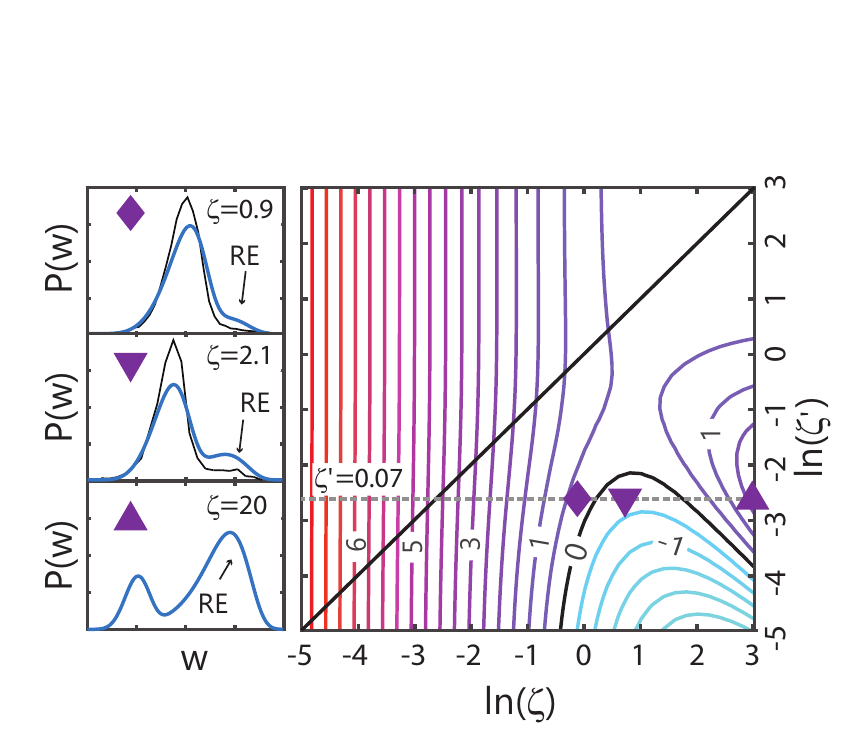}
	\caption{Contour plot of average work $\langle w \rangle_{\text c}^{\zeta'}$ as a function of $\zeta,\zeta'$. Negative work values are found for  small $\zeta'$ and large $\zeta$. Side panel: Work probability distributions for three sets of $\zeta,\zeta'$, marked in main panel. Trajectories with refolding events (RE) contribute to a shoulder at high work values, with height increasing with increasing $\zeta$. For two uppermost panels, the corresponding work probability distributions obtained from the dynamical simulations are shown (black, thin lines).}
\label{z1z2}
\end{figure}
The work probability distribution as well as the average work are obtained numerically, similarly to the idealized case \cite{SI}. Three representative probability distributions, for different $\zeta'$, are shown in Fig.~\ref{z1z2}. The common feature is that the distribution becomes bimodal, with an additional peak at positive work values developing due to the finite probability of refolding events $\text U \rightarrow \text F$ during the drive, with $\zeta'$, after the first unfolding. The average work $\langle w \rangle_{\text c}^{\zeta'}$ as a function of $\zeta,\zeta'$, shown in Fig.~\ref{z1z2}, is modified accordingly; any refolding after the first unfolding event will increase the work performed on the system. In fact, the average work can be written as  $\langle w \rangle_{\text c}^{\zeta'}=\langle w \rangle_{\text c}+\langle w \rangle^{\zeta'}$, a sum of the work performed under the continuous FB protocol with $\zeta'\rightarrow 0$, Eq.~(\ref{w_cM}), and the positive work, $\langle w \rangle^{\zeta'}>0$, due to the refolding events after the first transition. These results are compared to Monte Carlo simulations of SMFS of DNA hairpin folding experiments (see \cite{SI} for details), known to reproduce experimental results very well \cite{Alemany2014,AlemanyThesis}. The obtained work probability distributions, shown in the side panel of Fig. \ref{z1z2}, display the same overall features as the extended continuous FB model, including the average work extracted as well as clear signatures of the refolding events. 

{\em Conclusions ---} We have analyzed work extraction in a two-state model of a single molecule folding experiment, increasing our understanding of information-to-work conversion under discrete and continuous feedback and providing key guidance for future experiments.

\begin{acknowledgments}
R.S, J.J, P.P.P., P.S. were supported by the Swedish Research Council. The research leading to these results has received funding from the European Union's Seventh Framework program (FP7/2007-2013) under grant agreement No. 308850 (project acronym INFERNOS) and the Swedish Research Council project Nos. 2015-03824 and 2015-0612. M.R. and F.R. acknowledge support from European Union's Horizon 2020 Grant No. 687089, Spanish Research Council Grant FIS2016-80458-P and ICREA Academia Prize 2013. P.P.P. acknowledges funding from the European Union's Horizon 2020 research and innovation programme under the Marie Sk{\l}odowska-Curie Grant Agreement No. 796700.
\end{acknowledgments}

\bibliography{LicentiateP1}	

\begin{thebibliography}{66}%
\makeatletter
\providecommand \@ifxundefined [1]{%
 \@ifx{#1\undefined}
}%
\providecommand \@ifnum [1]{%
 \ifnum #1\expandafter \@firstoftwo
 \else \expandafter \@secondoftwo
 \fi
}%
\providecommand \@ifx [1]{%
 \ifx #1\expandafter \@firstoftwo
 \else \expandafter \@secondoftwo
 \fi
}%
\providecommand \natexlab [1]{#1}%
\providecommand \enquote  [1]{``#1''}%
\providecommand \bibnamefont  [1]{#1}%
\providecommand \bibfnamefont [1]{#1}%
\providecommand \citenamefont [1]{#1}%
\providecommand \href@noop [0]{\@secondoftwo}%
\providecommand \href [0]{\begingroup \@sanitize@url \@href}%
\providecommand \@href[1]{\@@startlink{#1}\@@href}%
\providecommand \@@href[1]{\endgroup#1\@@endlink}%
\providecommand \@sanitize@url [0]{\catcode `\\12\catcode `\$12\catcode
  `\&12\catcode `\#12\catcode `\^12\catcode `\_12\catcode `\%12\relax}%
\providecommand \@@startlink[1]{}%
\providecommand \@@endlink[0]{}%
\providecommand \url  [0]{\begingroup\@sanitize@url \@url }%
\providecommand \@url [1]{\endgroup\@href {#1}{\urlprefix }}%
\providecommand \urlprefix  [0]{URL }%
\providecommand \Eprint [0]{\href }%
\providecommand \doibase [0]{https://doi.org/}%
\providecommand \selectlanguage [0]{\@gobble}%
\providecommand \bibinfo  [0]{\@secondoftwo}%
\providecommand \bibfield  [0]{\@secondoftwo}%
\providecommand \translation [1]{[#1]}%
\providecommand \BibitemOpen [0]{}%
\providecommand \bibitemStop [0]{}%
\providecommand \bibitemNoStop [0]{.\EOS\space}%
\providecommand \EOS [0]{\spacefactor3000\relax}%
\providecommand \BibitemShut  [1]{\csname bibitem#1\endcsname}%
\let\auto@bib@innerbib\@empty
\bibitem [{\citenamefont {Harris}\ and\ \citenamefont
  {Sch\"utz}(2007)}]{Harris2007}%
  \BibitemOpen
  \bibfield  {author} {\bibinfo {author} {\bibfnamefont {R.~J.}\ \bibnamefont
  {Harris}}\ and\ \bibinfo {author} {\bibfnamefont {G.~M.}\ \bibnamefont
  {Sch\"utz}},\ }\bibfield  {title} {\bibinfo {title} {Fluctuation theorems for
  stochastic dynamics},\ }\href
  {http://stacks.iop.org/1742-5468/2007/i=07/a=P07020} {\bibfield  {journal}
  {\bibinfo  {journal} {J. Stat. Mech. Theor. Exp.}\ }\textbf {\bibinfo
  {volume} {2007}},\ \bibinfo {pages} {P07020} (\bibinfo {year}
  {2007})}\BibitemShut {NoStop}%
\bibitem [{\citenamefont {Esposito}\ \emph {et~al.}(2009)\citenamefont
  {Esposito}, \citenamefont {Harbola},\ and\ \citenamefont
  {Mukamel}}]{Esposito2009rmp}%
  \BibitemOpen
  \bibfield  {author} {\bibinfo {author} {\bibfnamefont {M.}~\bibnamefont
  {Esposito}}, \bibinfo {author} {\bibfnamefont {U.}~\bibnamefont {Harbola}},\
  and\ \bibinfo {author} {\bibfnamefont {S.}~\bibnamefont {Mukamel}},\
  }\bibfield  {title} {\bibinfo {title} {Nonequilibrium fluctuations,
  fluctuation theorems, and counting statistics in quantum systems},\ }\href
  {https://doi.org/10.1103/RevModPhys.81.1665} {\bibfield  {journal} {\bibinfo
  {journal} {Rev. Mod. Phys.}\ }\textbf {\bibinfo {volume} {81}},\ \bibinfo
  {pages} {1665} (\bibinfo {year} {2009})}\BibitemShut {NoStop}%
\bibitem [{\citenamefont {Jarzynski}(2011)}]{Jarzynski2011}%
  \BibitemOpen
  \bibfield  {author} {\bibinfo {author} {\bibfnamefont {C.}~\bibnamefont
  {Jarzynski}},\ }\bibfield  {title} {\bibinfo {title} {{Equalities and
  Inequalities: Irreversibility and the Second Law of Thermodynamics at the
  Nanoscale}},\ }\href
  {https://doi.org/10.1146/annurev-conmatphys-062910-140506} {\bibfield
  {journal} {\bibinfo  {journal} {Annu. Rev. Condens. Matter Phys.}\ }\textbf
  {\bibinfo {volume} {2}},\ \bibinfo {pages} {329} (\bibinfo {year}
  {2011})}\BibitemShut {NoStop}%
\bibitem [{\citenamefont {Seifert}(2012)}]{Seifert2012c}%
  \BibitemOpen
  \bibfield  {author} {\bibinfo {author} {\bibfnamefont {U.}~\bibnamefont
  {Seifert}},\ }\bibfield  {title} {\bibinfo {title} {{Stochastic
  thermodynamics, fluctuation theorems and molecular machines}},\ }\href
  {https://doi.org/10.1088/0034-4885/75/12/126001} {\bibfield  {journal}
  {\bibinfo  {journal} {Rep. Prog. Phys.}\ }\textbf {\bibinfo {volume} {75}},\
  \bibinfo {pages} {126001} (\bibinfo {year} {2012})}\BibitemShut {NoStop}%
\bibitem [{\citenamefont {{Malek Mansour}}\ and\ \citenamefont
  {Baras}(2017)}]{Mansour2017}%
  \BibitemOpen
  \bibfield  {author} {\bibinfo {author} {\bibfnamefont {M.}~\bibnamefont
  {{Malek Mansour}}}\ and\ \bibinfo {author} {\bibfnamefont {F.}~\bibnamefont
  {Baras}},\ }\bibfield  {title} {\bibinfo {title} {{Fluctuation theorem: A
  critical review}},\ }\href {https://doi.org/10.1063/1.4986600} {\bibfield
  {journal} {\bibinfo  {journal} {Chaos}\ }\textbf {\bibinfo {volume} {27}},\
  \bibinfo {pages} {104609} (\bibinfo {year} {2017})}\BibitemShut {NoStop}%
\bibitem [{\citenamefont {Campisi}\ \emph {et~al.}(2011)\citenamefont
  {Campisi}, \citenamefont {H\"anggi},\ and\ \citenamefont
  {Talkner}}]{campisi:2011}%
  \BibitemOpen
  \bibfield  {author} {\bibinfo {author} {\bibfnamefont {M.}~\bibnamefont
  {Campisi}}, \bibinfo {author} {\bibfnamefont {P.}~\bibnamefont {H\"anggi}},\
  and\ \bibinfo {author} {\bibfnamefont {P.}~\bibnamefont {Talkner}},\
  }\bibfield  {title} {\bibinfo {title} {Colloquium: Quantum fluctuation
  relations: Foundations and applications},\ }\href
  {https://doi.org/10.1103/RevModPhys.83.771} {\bibfield  {journal} {\bibinfo
  {journal} {Rev. Mod. Phys.}\ }\textbf {\bibinfo {volume} {83}},\ \bibinfo
  {pages} {771} (\bibinfo {year} {2011})}\BibitemShut {NoStop}%
\bibitem [{\citenamefont {Crooks}(1998)}]{Crooks1998}%
  \BibitemOpen
  \bibfield  {author} {\bibinfo {author} {\bibfnamefont {G.~E.}\ \bibnamefont
  {Crooks}},\ }\bibfield  {title} {\bibinfo {title} {{Nonequilibrium
  Measurements of Free Energy Differences for Microscopically Reversible
  Markovian Systems}},\ }\href {https://doi.org/10.1023/A:1023208217925}
  {\bibfield  {journal} {\bibinfo  {journal} {J. Stat. Phys.}\ }\textbf
  {\bibinfo {volume} {90}},\ \bibinfo {pages} {1481} (\bibinfo {year}
  {1998})}\BibitemShut {NoStop}%
\bibitem [{\citenamefont {Crooks}(1999)}]{Crooks1999}%
  \BibitemOpen
  \bibfield  {author} {\bibinfo {author} {\bibfnamefont {G.~E.}\ \bibnamefont
  {Crooks}},\ }\bibfield  {title} {\bibinfo {title} {{The Entropy Production
  Fluctuation Theorem and the Nonequilibrium Work Relation for Free Energy
  Differences}},\ }\href
  {http://arxiv.org/abs/cond-mat/9901352{\%}5Cnhttp://dx.doi.org/10.1103/PhysRevE.60.2721}
  {\bibfield  {journal} {\bibinfo  {journal} {Phys. Rev. E}\ }\textbf {\bibinfo
  {volume} {60}},\ \bibinfo {pages} {2721} (\bibinfo {year}
  {1999})}\BibitemShut {NoStop}%
\bibitem [{\citenamefont {Crooks}(2000)}]{Crooks2000}%
  \BibitemOpen
  \bibfield  {author} {\bibinfo {author} {\bibfnamefont {G.~E.}\ \bibnamefont
  {Crooks}},\ }\bibfield  {title} {\bibinfo {title} {{Path-ensemble averages in
  systems driven far from equilibrium}},\ }\href
  {http://link.aps.org/doi/10.1103/PhysRevE.61.2361{\%}5Cnhttp://pre.aps.org/pdf/PRE/v61/i3/p2361{\%}7B{\_}{\%}7D1
  http://link.aps.org/doi/10.1103/PhysRevE.61.2361{\%}5Cnhttp://pre.aps.org/pdf/PRE/v61/i3/p2361{\_}1}
  {\bibfield  {journal} {\bibinfo  {journal} {Phys. Rev. E}\ }\textbf {\bibinfo
  {volume} {61}},\ \bibinfo {pages} {2361} (\bibinfo {year}
  {2000})}\BibitemShut {NoStop}%
\bibitem [{\citenamefont {Jarzynski}(1997{\natexlab{a}})}]{Jarzynski1997a}%
  \BibitemOpen
  \bibfield  {author} {\bibinfo {author} {\bibfnamefont {C.}~\bibnamefont
  {Jarzynski}},\ }\bibfield  {title} {\bibinfo {title} {{Nonequilibrium
  Equality for Free Energy Differences}},\ }\href
  {http://link.aps.org/doi/10.1103/PhysRevLett.78.2690} {\bibfield  {journal}
  {\bibinfo  {journal} {Phys. Rev. Lett.}\ }\textbf {\bibinfo {volume} {78}},\
  \bibinfo {pages} {2690} (\bibinfo {year} {1997}{\natexlab{a}})}\BibitemShut
  {NoStop}%
\bibitem [{\citenamefont {Jarzynski}(1997{\natexlab{b}})}]{Jarzynski1997}%
  \BibitemOpen
  \bibfield  {author} {\bibinfo {author} {\bibfnamefont {C.}~\bibnamefont
  {Jarzynski}},\ }\bibfield  {title} {\bibinfo {title} {{Equilibrium
  free-energy differences from nonequilibrium measurements: A master-equation
  approach}},\ }\href {http://link.aps.org/doi/10.1103/PhysRevE.56.5018}
  {\bibfield  {journal} {\bibinfo  {journal} {Phys. Rev. E}\ }\textbf {\bibinfo
  {volume} {56}},\ \bibinfo {pages} {5018} (\bibinfo {year}
  {1997}{\natexlab{b}})}\BibitemShut {NoStop}%
\bibitem [{\citenamefont {Sagawa}\ and\ \citenamefont
  {Ueda}(2008)}]{Sagawa2008}%
  \BibitemOpen
  \bibfield  {author} {\bibinfo {author} {\bibfnamefont {T.}~\bibnamefont
  {Sagawa}}\ and\ \bibinfo {author} {\bibfnamefont {M.}~\bibnamefont {Ueda}},\
  }\bibfield  {title} {\bibinfo {title} {Second law of thermodynamics with
  discrete quantum feedback control},\ }\href
  {https://link.aps.org/doi/10.1103/PhysRevLett.100.080403} {\bibfield
  {journal} {\bibinfo  {journal} {Phys. Rev. Lett.}\ }\textbf {\bibinfo
  {volume} {100}},\ \bibinfo {pages} {080403} (\bibinfo {year}
  {2008})}\BibitemShut {NoStop}%
\bibitem [{\citenamefont {Cao}\ and\ \citenamefont
  {Feito}(2009)}]{CaoFeito2009}%
  \BibitemOpen
  \bibfield  {author} {\bibinfo {author} {\bibfnamefont {F.}~\bibnamefont
  {Cao}}\ and\ \bibinfo {author} {\bibfnamefont {M.}~\bibnamefont {Feito}},\
  }\bibfield  {title} {\bibinfo {title} {{Thermodynamics of feedback controlled
  systems}},\ }\href {http://link.aps.org/doi/10.1103/PhysRevE.79.041118}
  {\bibfield  {journal} {\bibinfo  {journal} {Phys. Rev. E}\ }\textbf {\bibinfo
  {volume} {79}},\ \bibinfo {pages} {041118} (\bibinfo {year}
  {2009})}\BibitemShut {NoStop}%
\bibitem [{\citenamefont {Sagawa}\ and\ \citenamefont
  {Ueda}(2010)}]{Sagawa2010}%
  \BibitemOpen
  \bibfield  {author} {\bibinfo {author} {\bibfnamefont {T.}~\bibnamefont
  {Sagawa}}\ and\ \bibinfo {author} {\bibfnamefont {M.}~\bibnamefont {Ueda}},\
  }\bibfield  {title} {\bibinfo {title} {{Generalized {J}arzynski equality
  under nonequilibrium feedback control}},\ }\href
  {https://doi.org/10.1103/PhysRevLett.104.090602} {\bibfield  {journal}
  {\bibinfo  {journal} {Phys. Rev. Lett.}\ }\textbf {\bibinfo {volume} {104}},\
  \bibinfo {pages} {090602} (\bibinfo {year} {2010})}\BibitemShut {NoStop}%
\bibitem [{\citenamefont {Horowitz}\ and\ \citenamefont
  {Vaikuntanathan}(2010)}]{Horowitz2010}%
  \BibitemOpen
  \bibfield  {author} {\bibinfo {author} {\bibfnamefont {J.~M.}\ \bibnamefont
  {Horowitz}}\ and\ \bibinfo {author} {\bibfnamefont {S.}~\bibnamefont
  {Vaikuntanathan}},\ }\bibfield  {title} {\bibinfo {title} {{Nonequilibrium
  detailed fluctuation theorem for repeated discrete feedback}},\ }\href
  {https://doi.org/10.1103/PhysRevE.82.061120} {\bibfield  {journal} {\bibinfo
  {journal} {Phys. Rev. E}\ }\textbf {\bibinfo {volume} {82}},\ \bibinfo
  {pages} {061120} (\bibinfo {year} {2010})}\BibitemShut {NoStop}%
\bibitem [{\citenamefont {Ponmurugan}(2010)}]{Ponmurugan2010}%
  \BibitemOpen
  \bibfield  {author} {\bibinfo {author} {\bibfnamefont {M.}~\bibnamefont
  {Ponmurugan}},\ }\bibfield  {title} {\bibinfo {title} {{Generalized detailed
  fluctuation theorem under nonequilibrium feedback control}},\ }\href
  {https://doi.org/10.1103/PhysRevE.82.031129} {\bibfield  {journal} {\bibinfo
  {journal} {Phys. Rev. E}\ }\textbf {\bibinfo {volume} {82}},\ \bibinfo
  {pages} {031129} (\bibinfo {year} {2010})}\BibitemShut {NoStop}%
\bibitem [{\citenamefont {Sagawa}(2011)}]{Sagawa2011a}%
  \BibitemOpen
  \bibfield  {author} {\bibinfo {author} {\bibfnamefont {T.}~\bibnamefont
  {Sagawa}},\ }\bibfield  {title} {\bibinfo {title} {{Hamiltonian Derivations
  of the Generalized Jarzynski Equalities under Feedback Control}},\ }\href
  {https://doi.org/10.1088/1742-6596/297/1/012015} {\bibfield  {journal}
  {\bibinfo  {journal} {J. Phys. Conf. Ser.}\ }\textbf {\bibinfo {volume}
  {297}},\ \bibinfo {pages} {2015} (\bibinfo {year} {2011})}\BibitemShut
  {NoStop}%
\bibitem [{\citenamefont {Sagawa}\ and\ \citenamefont
  {Ueda}(2012)}]{Sagawa2012}%
  \BibitemOpen
  \bibfield  {author} {\bibinfo {author} {\bibfnamefont {T.}~\bibnamefont
  {Sagawa}}\ and\ \bibinfo {author} {\bibfnamefont {M.}~\bibnamefont {Ueda}},\
  }\bibfield  {title} {\bibinfo {title} {{Nonequilibrium thermodynamics of
  feedback control}},\ }\href {https://doi.org/10.1103/PhysRevE.85.021104}
  {\bibfield  {journal} {\bibinfo  {journal} {Phys. Rev. E}\ }\textbf {\bibinfo
  {volume} {85}},\ \bibinfo {pages} {021104} (\bibinfo {year}
  {2012})}\BibitemShut {NoStop}%
\bibitem [{\citenamefont {Lahiri}\ \emph {et~al.}(2012)\citenamefont {Lahiri},
  \citenamefont {Rana},\ and\ \citenamefont {Jayannavar}}]{Lahiri2012}%
  \BibitemOpen
  \bibfield  {author} {\bibinfo {author} {\bibfnamefont {S.}~\bibnamefont
  {Lahiri}}, \bibinfo {author} {\bibfnamefont {S.}~\bibnamefont {Rana}},\ and\
  \bibinfo {author} {\bibfnamefont {A.~M.}\ \bibnamefont {Jayannavar}},\
  }\bibfield  {title} {\bibinfo {title} {{Fluctuation theorems in the presence
  of information gain and feedback}},\ }\href
  {https://doi.org/10.1088/1751-8113/45/6/065002} {\bibfield  {journal}
  {\bibinfo  {journal} {J. Phys. A: Math. Theor.}\ }\textbf {\bibinfo {volume}
  {45}},\ \bibinfo {pages} {065002} (\bibinfo {year} {2012})}\BibitemShut
  {NoStop}%
\bibitem [{\citenamefont {Abreu}\ and\ \citenamefont
  {Seifert}(2012)}]{Abreu2012}%
  \BibitemOpen
  \bibfield  {author} {\bibinfo {author} {\bibfnamefont {D.}~\bibnamefont
  {Abreu}}\ and\ \bibinfo {author} {\bibfnamefont {U.}~\bibnamefont
  {Seifert}},\ }\bibfield  {title} {\bibinfo {title} {{Thermodynamics of
  Genuine Nonequilibrium States under Feedback Control}},\ }\href
  {https://doi.org/10.1103/PhysRevLett.108.030601} {\bibfield  {journal}
  {\bibinfo  {journal} {Phys. Rev. Lett.}\ }\textbf {\bibinfo {volume} {108}},\
  \bibinfo {pages} {30601} (\bibinfo {year} {2012})}\BibitemShut {NoStop}%
\bibitem [{\citenamefont {Ashida}\ \emph {et~al.}(2014)\citenamefont {Ashida},
  \citenamefont {Funo},\ and\ \citenamefont {Ueda}}]{Ashida2014}%
  \BibitemOpen
  \bibfield  {author} {\bibinfo {author} {\bibfnamefont {Y.}~\bibnamefont
  {Ashida}}, \bibinfo {author} {\bibfnamefont {Y.}~\bibnamefont {Funo},
  \bibfnamefont {K.and~Murashita}},\ and\ \bibinfo {author} {\bibfnamefont
  {M.}~\bibnamefont {Ueda}},\ }\bibfield  {title} {\bibinfo {title} {{General
  achievable bound of extractable work under feedback control}},\ }\href
  {https://doi.org/10.1103/PhysRevE.90.052125} {\bibfield  {journal} {\bibinfo
  {journal} {Phys. Rev. E}\ }\textbf {\bibinfo {volume} {90}},\ \bibinfo
  {pages} {052125} (\bibinfo {year} {2014})}\BibitemShut {NoStop}%
\bibitem [{\citenamefont {Horowitz}\ and\ \citenamefont
  {Sandberg}(2014)}]{Horowitz2014}%
  \BibitemOpen
  \bibfield  {author} {\bibinfo {author} {\bibfnamefont {J.~M.}\ \bibnamefont
  {Horowitz}}\ and\ \bibinfo {author} {\bibfnamefont {H.}~\bibnamefont
  {Sandberg}},\ }\bibfield  {title} {\bibinfo {title} {Second-law-like
  inequalities with information and their interpretations},\ }\href
  {http://stacks.iop.org/1367-2630/16/i=12/a=125007} {\bibfield  {journal}
  {\bibinfo  {journal} {New J. Phys.}\ }\textbf {\bibinfo {volume} {16}},\
  \bibinfo {pages} {125007} (\bibinfo {year} {2014})}\BibitemShut {NoStop}%
\bibitem [{\citenamefont {Horowitz}\ and\ \citenamefont
  {Esposito}(2014)}]{Horowitz2014prx}%
  \BibitemOpen
  \bibfield  {author} {\bibinfo {author} {\bibfnamefont {J.~M.}\ \bibnamefont
  {Horowitz}}\ and\ \bibinfo {author} {\bibfnamefont {M.}~\bibnamefont
  {Esposito}},\ }\bibfield  {title} {\bibinfo {title} {Thermodynamics with
  continuous information flow},\ }\href
  {https://link.aps.org/doi/10.1103/PhysRevX.4.031015} {\bibfield  {journal}
  {\bibinfo  {journal} {Phys. Rev. X}\ }\textbf {\bibinfo {volume} {4}},\
  \bibinfo {pages} {031015} (\bibinfo {year} {2014})}\BibitemShut {NoStop}%
\bibitem [{\citenamefont {W\"achtler}\ \emph {et~al.}(2016)\citenamefont
  {W\"achtler}, \citenamefont {Strasberg},\ and\ \citenamefont
  {Brandes}}]{Wachtler2016}%
  \BibitemOpen
  \bibfield  {author} {\bibinfo {author} {\bibfnamefont {C.~W.}\ \bibnamefont
  {W\"achtler}}, \bibinfo {author} {\bibfnamefont {P.}~\bibnamefont
  {Strasberg}},\ and\ \bibinfo {author} {\bibfnamefont {T.}~\bibnamefont
  {Brandes}},\ }\bibfield  {title} {\bibinfo {title} {Stochastic thermodynamics
  based on incomplete information: generalized {J}arzynski equality with
  measurement errors with or without feedback},\ }\href
  {http://stacks.iop.org/1367-2630/18/i=11/a=113042} {\bibfield  {journal}
  {\bibinfo  {journal} {New J. Phys.}\ }\textbf {\bibinfo {volume} {18}},\
  \bibinfo {pages} {113042} (\bibinfo {year} {2016})}\BibitemShut {NoStop}%
\bibitem [{\citenamefont {Potts}\ and\ \citenamefont
  {Samuelsson}(2018)}]{potts:2018}%
  \BibitemOpen
  \bibfield  {author} {\bibinfo {author} {\bibfnamefont {P.~P.}\ \bibnamefont
  {Potts}}\ and\ \bibinfo {author} {\bibfnamefont {P.}~\bibnamefont
  {Samuelsson}},\ }\bibfield  {title} {\bibinfo {title} {Detailed fluctuation
  relation for arbitrary measurement and feedback schemes},\ }\href
  {https://doi.org/10.1103/PhysRevLett.121.210603} {\bibfield  {journal}
  {\bibinfo  {journal} {Phys. Rev. Lett.}\ }\textbf {\bibinfo {volume} {121}},\
  \bibinfo {pages} {210603} (\bibinfo {year} {2018})}\BibitemShut {NoStop}%
\bibitem [{\citenamefont {Toyabe}\ \emph {et~al.}(2010)\citenamefont {Toyabe},
  \citenamefont {Sagawa}, \citenamefont {Ueda}, \citenamefont {Muneyuki},\ and\
  \citenamefont {Sano}}]{Toyabe2010}%
  \BibitemOpen
  \bibfield  {author} {\bibinfo {author} {\bibfnamefont {S.}~\bibnamefont
  {Toyabe}}, \bibinfo {author} {\bibfnamefont {T.}~\bibnamefont {Sagawa}},
  \bibinfo {author} {\bibfnamefont {M.}~\bibnamefont {Ueda}}, \bibinfo {author}
  {\bibfnamefont {E.}~\bibnamefont {Muneyuki}},\ and\ \bibinfo {author}
  {\bibfnamefont {M.}~\bibnamefont {Sano}},\ }\bibfield  {title} {\bibinfo
  {title} {{Experimental demonstration of information-to-energy conversion and
  validation of the generalized Jarzynski equality}},\ }\href
  {https://doi.org/10.1038/nphys1821} {\bibfield  {journal} {\bibinfo
  {journal} {Nat. Phys.}\ }\textbf {\bibinfo {volume} {6}},\ \bibinfo {pages}
  {988} (\bibinfo {year} {2010})}\BibitemShut {NoStop}%
\bibitem [{\citenamefont {Fujitani}\ and\ \citenamefont
  {Suzuki}(2010)}]{Fujitani2010}%
  \BibitemOpen
  \bibfield  {author} {\bibinfo {author} {\bibfnamefont {Y.}~\bibnamefont
  {Fujitani}}\ and\ \bibinfo {author} {\bibfnamefont {H.}~\bibnamefont
  {Suzuki}},\ }\bibfield  {title} {\bibinfo {title} {{Jarzynski equality
  modified in the linear feedback system}},\ }\href
  {https://doi.org/10.1143/JPSJ.79.104003} {\bibfield  {journal} {\bibinfo
  {journal} {J. Phys. Soc. Jpn.}\ }\textbf {\bibinfo {volume} {79}},\ \bibinfo
  {pages} {104003} (\bibinfo {year} {2010})}\BibitemShut {NoStop}%
\bibitem [{\citenamefont {Landauer}(1961)}]{Landauer1961}%
  \BibitemOpen
  \bibfield  {author} {\bibinfo {author} {\bibfnamefont {R.}~\bibnamefont
  {Landauer}},\ }\bibfield  {title} {\bibinfo {title} {{Irreversibility and
  Heat Generation in the Computing Process}},\ }\href
  {https://doi.org/10.1147/rd.53.0183} {\bibfield  {journal} {\bibinfo
  {journal} {IBM J. Res. Dev.}\ }\textbf {\bibinfo {volume} {5}},\ \bibinfo
  {pages} {183} (\bibinfo {year} {1961})}\BibitemShut {NoStop}%
\bibitem [{\citenamefont {B{\'{e}}rut}\ \emph {et~al.}(2012)\citenamefont
  {B{\'{e}}rut}, \citenamefont {Arakelyan}, \citenamefont {Petrosyan},
  \citenamefont {Ciliberto}, \citenamefont {Dillenschneider},\ and\
  \citenamefont {Lutz}}]{Berut2012}%
  \BibitemOpen
  \bibfield  {author} {\bibinfo {author} {\bibfnamefont {A.}~\bibnamefont
  {B{\'{e}}rut}}, \bibinfo {author} {\bibfnamefont {A.}~\bibnamefont
  {Arakelyan}}, \bibinfo {author} {\bibfnamefont {A.}~\bibnamefont
  {Petrosyan}}, \bibinfo {author} {\bibfnamefont {S.}~\bibnamefont
  {Ciliberto}}, \bibinfo {author} {\bibfnamefont {R.}~\bibnamefont
  {Dillenschneider}},\ and\ \bibinfo {author} {\bibfnamefont {E.}~\bibnamefont
  {Lutz}},\ }\bibfield  {title} {\bibinfo {title} {{Experimental verification
  of Landauer's principle linking information and thermodynamics.}},\ }\href
  {https://doi.org/10.1038/nature10872} {\bibfield  {journal} {\bibinfo
  {journal} {Nature}\ }\textbf {\bibinfo {volume} {483}},\ \bibinfo {pages}
  {187} (\bibinfo {year} {2012})}\BibitemShut {NoStop}%
\bibitem [{\citenamefont {Jun}\ \emph {et~al.}(2014)\citenamefont {Jun},
  \citenamefont {Gavrilov},\ and\ \citenamefont {Bechhoefer}}]{Jun2014}%
  \BibitemOpen
  \bibfield  {author} {\bibinfo {author} {\bibfnamefont {.}~\bibnamefont
  {Jun}}, \bibinfo {author} {\bibfnamefont {M.}~\bibnamefont {Gavrilov}},\ and\
  \bibinfo {author} {\bibfnamefont {J.}~\bibnamefont {Bechhoefer}},\ }\bibfield
   {title} {\bibinfo {title} {{High-precision test of Landauer's principle in a
  feedback trap}},\ }\href {https://doi.org/10.1103/PhysRevLett.113.190601}
  {\bibfield  {journal} {\bibinfo  {journal} {Phys. Rev. Lett.}\ }\textbf
  {\bibinfo {volume} {113}},\ \bibinfo {pages} {190601} (\bibinfo {year}
  {2014})}\BibitemShut {NoStop}%
\bibitem [{\citenamefont {Maxwell}(1871)}]{Maxwell1871}%
  \BibitemOpen
  \bibfield  {author} {\bibinfo {author} {\bibfnamefont {J.~C.}\ \bibnamefont
  {Maxwell}},\ }\href@noop {} {\emph {\bibinfo {title} {Theory of Heat}}}\
  (\bibinfo  {publisher} {Longmans, Green, and Co.},\ \bibinfo {year}
  {1871})\BibitemShut {NoStop}%
\bibitem [{\citenamefont {Szilard}(1929)}]{Szilard1929}%
  \BibitemOpen
  \bibfield  {author} {\bibinfo {author} {\bibfnamefont {L.}~\bibnamefont
  {Szilard}},\ }\bibfield  {title} {\bibinfo {title} {{{\"{U}}ber die
  Entropieverminderung in einem thermodynamischen System bei Eingriffen
  intelligenter Wesen}},\ }\href {https://doi.org/10.1007/BF01341281}
  {\bibfield  {journal} {\bibinfo  {journal} {Z. Phys.}\ }\textbf {\bibinfo
  {volume} {53}},\ \bibinfo {pages} {840} (\bibinfo {year} {1929})}\BibitemShut
  {NoStop}%
\bibitem [{\citenamefont {Koski}\ \emph {et~al.}(2014)\citenamefont {Koski},
  \citenamefont {Maisi}, \citenamefont {Pekola},\ and\ \citenamefont
  {Averin}}]{Koski2014}%
  \BibitemOpen
  \bibfield  {author} {\bibinfo {author} {\bibfnamefont {J.~V.~.}\ \bibnamefont
  {Koski}}, \bibinfo {author} {\bibfnamefont {V.~F.}\ \bibnamefont {Maisi}},
  \bibinfo {author} {\bibfnamefont {J.~P.}\ \bibnamefont {Pekola}},\ and\
  \bibinfo {author} {\bibfnamefont {D.~V.}\ \bibnamefont {Averin}},\ }\bibfield
   {title} {\bibinfo {title} {{Experimental realization of a Szilard engine
  with a single electron.}},\ }\href {http://www.pnas.org/content/111/38/13786}
  {\bibfield  {journal} {\bibinfo  {journal} {Proc. Natl. Acad. Sci.}\ }\textbf
  {\bibinfo {volume} {111}},\ \bibinfo {pages} {13786} (\bibinfo {year}
  {2014})}\BibitemShut {NoStop}%
\bibitem [{\citenamefont {Koski}\ \emph {et~al.}(2015)\citenamefont {Koski},
  \citenamefont {Kutvonen}, \citenamefont {Khaymovich}, \citenamefont
  {Ala-Nissila},\ and\ \citenamefont {Pekola}}]{Koski2015}%
  \BibitemOpen
  \bibfield  {author} {\bibinfo {author} {\bibfnamefont {J.~V.}\ \bibnamefont
  {Koski}}, \bibinfo {author} {\bibfnamefont {A.}~\bibnamefont {Kutvonen}},
  \bibinfo {author} {\bibfnamefont {I.~M.}\ \bibnamefont {Khaymovich}},
  \bibinfo {author} {\bibfnamefont {T.}~\bibnamefont {Ala-Nissila}},\ and\
  \bibinfo {author} {\bibfnamefont {J.~P.}\ \bibnamefont {Pekola}},\ }\bibfield
   {title} {\bibinfo {title} {{On-Chip Maxwell's Demon as an
  Information-Powered Refrigerator}},\ }\href
  {https://doi.org/10.1103/PhysRevLett.115.260602} {\bibfield  {journal}
  {\bibinfo  {journal} {Phys. Rev. Lett.}\ }\textbf {\bibinfo {volume} {115}},\
  \bibinfo {pages} {260602} (\bibinfo {year} {2015})}\BibitemShut {NoStop}%
\bibitem [{\citenamefont {Chida}\ \emph {et~al.}(2017)\citenamefont {Chida},
  \citenamefont {Desai}, \citenamefont {Nishiguchi},\ and\ \citenamefont
  {Fujiwara}}]{Chida2017}%
  \BibitemOpen
  \bibfield  {author} {\bibinfo {author} {\bibfnamefont {K.}~\bibnamefont
  {Chida}}, \bibinfo {author} {\bibfnamefont {S.}~\bibnamefont {Desai}},
  \bibinfo {author} {\bibfnamefont {K.}~\bibnamefont {Nishiguchi}},\ and\
  \bibinfo {author} {\bibfnamefont {A.}~\bibnamefont {Fujiwara}},\ }\bibfield
  {title} {\bibinfo {title} {Power generator driven by {M}axwell’s demon},\
  }\href {http://dx.doi.org/10.1038/ncomms15301} {\bibfield  {journal}
  {\bibinfo  {journal} {Nat. Commun.}\ }\textbf {\bibinfo {volume} {8}},\
  \bibinfo {pages} {15310} (\bibinfo {year} {2017})}\BibitemShut {NoStop}%
\bibitem [{\citenamefont {Cottet}\ \emph {et~al.}(2017)\citenamefont {Cottet},
  \citenamefont {Jezouin}, \citenamefont {Bretheau}, \citenamefont
  {Campagne-Ibarcq}, \citenamefont {Ficheux}, \citenamefont {Anders},
  \citenamefont {Auff{\`e}ves}, \citenamefont {Azouit}, \citenamefont
  {Rouchon},\ and\ \citenamefont {Huard}}]{Cottet2017}%
  \BibitemOpen
  \bibfield  {author} {\bibinfo {author} {\bibfnamefont {N.}~\bibnamefont
  {Cottet}}, \bibinfo {author} {\bibfnamefont {S.}~\bibnamefont {Jezouin}},
  \bibinfo {author} {\bibfnamefont {L.}~\bibnamefont {Bretheau}}, \bibinfo
  {author} {\bibfnamefont {P.}~\bibnamefont {Campagne-Ibarcq}}, \bibinfo
  {author} {\bibfnamefont {Q.}~\bibnamefont {Ficheux}}, \bibinfo {author}
  {\bibfnamefont {J.}~\bibnamefont {Anders}}, \bibinfo {author} {\bibfnamefont
  {A.}~\bibnamefont {Auff{\`e}ves}}, \bibinfo {author} {\bibfnamefont
  {R.}~\bibnamefont {Azouit}}, \bibinfo {author} {\bibfnamefont
  {P.}~\bibnamefont {Rouchon}},\ and\ \bibinfo {author} {\bibfnamefont
  {B.}~\bibnamefont {Huard}},\ }\bibfield  {title} {\bibinfo {title} {Observing
  a quantum {M}axwell demon at work},\ }\href
  {http://www.pnas.org/content/114/29/7561} {\bibfield  {journal} {\bibinfo
  {journal} {Proc. Natl. Acad. Sci. USA}\ }\textbf {\bibinfo {volume} {114}},\
  \bibinfo {pages} {7561} (\bibinfo {year} {2017})}\BibitemShut {NoStop}%
\bibitem [{\citenamefont {Masuyama}\ \emph {et~al.}(2018)\citenamefont
  {Masuyama}, \citenamefont {Funo}, \citenamefont {Murashita}, \citenamefont
  {Noguchi}, \citenamefont {Kono}, \citenamefont {Tabuchi}, \citenamefont
  {Yamazaki}, \citenamefont {Ueda},\ and\ \citenamefont
  {Nakamura}}]{Masuyama2018}%
  \BibitemOpen
  \bibfield  {author} {\bibinfo {author} {\bibfnamefont {Y.}~\bibnamefont
  {Masuyama}}, \bibinfo {author} {\bibfnamefont {K.}~\bibnamefont {Funo}},
  \bibinfo {author} {\bibfnamefont {Y.}~\bibnamefont {Murashita}}, \bibinfo
  {author} {\bibfnamefont {A.}~\bibnamefont {Noguchi}}, \bibinfo {author}
  {\bibfnamefont {S.}~\bibnamefont {Kono}}, \bibinfo {author} {\bibfnamefont
  {Y.}~\bibnamefont {Tabuchi}}, \bibinfo {author} {\bibfnamefont
  {R.}~\bibnamefont {Yamazaki}}, \bibinfo {author} {\bibfnamefont
  {M.}~\bibnamefont {Ueda}},\ and\ \bibinfo {author} {\bibfnamefont
  {Y.}~\bibnamefont {Nakamura}},\ }\bibfield  {title} {\bibinfo {title}
  {Information-to-work conversion by {M}axwell’s demon in a superconducting
  circuit quantum electrodynamical system},\ }\href
  {https://doi.org/10.1038/s41467-018-03686-y} {\bibfield  {journal} {\bibinfo
  {journal} {Nat. Commun.}\ }\textbf {\bibinfo {volume} {9}},\ \bibinfo {pages}
  {1291} (\bibinfo {year} {2018})}\BibitemShut {NoStop}%
\bibitem [{\citenamefont {Naghiloo}\ \emph {et~al.}(2018)\citenamefont
  {Naghiloo}, \citenamefont {Alonso}, \citenamefont {Romito}, \citenamefont
  {Lutz},\ and\ \citenamefont {Murch}}]{Naghiloo2018}%
  \BibitemOpen
  \bibfield  {author} {\bibinfo {author} {\bibfnamefont {M.}~\bibnamefont
  {Naghiloo}}, \bibinfo {author} {\bibfnamefont {J.~J.}\ \bibnamefont
  {Alonso}}, \bibinfo {author} {\bibfnamefont {A.}~\bibnamefont {Romito}},
  \bibinfo {author} {\bibfnamefont {E.}~\bibnamefont {Lutz}},\ and\ \bibinfo
  {author} {\bibfnamefont {K.~W.}\ \bibnamefont {Murch}},\ }\bibfield  {title}
  {\bibinfo {title} {Information gain and loss for a quantum maxwell's demon},\
  }\href {https://doi.org/10.1103/PhysRevLett.121.030604} {\bibfield  {journal}
  {\bibinfo  {journal} {Phys. Rev. Lett.}\ }\textbf {\bibinfo {volume} {121}},\
  \bibinfo {pages} {030604} (\bibinfo {year} {2018})}\BibitemShut {NoStop}%
\bibitem [{\citenamefont {Vidrighin}\ \emph {et~al.}(2016)\citenamefont
  {Vidrighin}, \citenamefont {Dahlsten}, \citenamefont {Barbieri},
  \citenamefont {Kim}, \citenamefont {Vedral},\ and\ \citenamefont
  {Walmsley}}]{Vidrighin2016}%
  \BibitemOpen
  \bibfield  {author} {\bibinfo {author} {\bibfnamefont {M.~D.}\ \bibnamefont
  {Vidrighin}}, \bibinfo {author} {\bibfnamefont {O.}~\bibnamefont {Dahlsten}},
  \bibinfo {author} {\bibfnamefont {M.}~\bibnamefont {Barbieri}}, \bibinfo
  {author} {\bibfnamefont {M.~S.}\ \bibnamefont {Kim}}, \bibinfo {author}
  {\bibfnamefont {V.}~\bibnamefont {Vedral}},\ and\ \bibinfo {author}
  {\bibfnamefont {I.~A.}\ \bibnamefont {Walmsley}},\ }\bibfield  {title}
  {\bibinfo {title} {{Photonic Maxwell's Demon}},\ }\href
  {http://link.aps.org/doi/10.1103/PhysRevLett.116.050401} {\bibfield
  {journal} {\bibinfo  {journal} {Phys. Rev. Lett.}\ }\textbf {\bibinfo
  {volume} {116}},\ \bibinfo {pages} {050401} (\bibinfo {year}
  {2016})}\BibitemShut {NoStop}%
\bibitem [{\citenamefont {Wang}\ \emph {et~al.}(2002)\citenamefont {Wang},
  \citenamefont {Sevick}, \citenamefont {Mittag}, \citenamefont {Searles},\
  and\ \citenamefont {Evans}}]{Wang2002a}%
  \BibitemOpen
  \bibfield  {author} {\bibinfo {author} {\bibfnamefont {G.~M.}\ \bibnamefont
  {Wang}}, \bibinfo {author} {\bibfnamefont {E.~M.}\ \bibnamefont {Sevick}},
  \bibinfo {author} {\bibfnamefont {E.}~\bibnamefont {Mittag}}, \bibinfo
  {author} {\bibfnamefont {D.~J.}\ \bibnamefont {Searles}},\ and\ \bibinfo
  {author} {\bibfnamefont {D.~J.}\ \bibnamefont {Evans}},\ }\bibfield  {title}
  {\bibinfo {title} {{Experimental demonstration of violations of the second
  law of thermodynamics for small systems and short time scales}},\ }\href
  {https://doi.org/10.1103/PhysRevLett.89.050601} {\bibfield  {journal}
  {\bibinfo  {journal} {Phys. Rev. Lett.}\ }\textbf {\bibinfo {volume} {89}},\
  \bibinfo {pages} {50601} (\bibinfo {year} {2002})}\BibitemShut {NoStop}%
\bibitem [{\citenamefont {Trepagnier}\ \emph {et~al.}(2004)\citenamefont
  {Trepagnier}, \citenamefont {Jarzynski}, \citenamefont {Ritort},
  \citenamefont {Crooks}, \citenamefont {Bustamante},\ and\ \citenamefont
  {Liphardt}}]{Trepagnier2004}%
  \BibitemOpen
  \bibfield  {author} {\bibinfo {author} {\bibfnamefont {E.~H.}\ \bibnamefont
  {Trepagnier}}, \bibinfo {author} {\bibfnamefont {C.}~\bibnamefont
  {Jarzynski}}, \bibinfo {author} {\bibfnamefont {F.}~\bibnamefont {Ritort}},
  \bibinfo {author} {\bibfnamefont {G.~E.}\ \bibnamefont {Crooks}}, \bibinfo
  {author} {\bibfnamefont {C.~J.}\ \bibnamefont {Bustamante}},\ and\ \bibinfo
  {author} {\bibfnamefont {J.}~\bibnamefont {Liphardt}},\ }\bibfield  {title}
  {\bibinfo {title} {{Experimental test of Hatano and Sasa's nonequilibrium
  steady-state equality}},\ }\href {https://doi.org/10.1073/pnas.0406405101}
  {\bibfield  {journal} {\bibinfo  {journal} {Proc. Natl. Acad. Sci.}\ }\textbf
  {\bibinfo {volume} {101}},\ \bibinfo {pages} {15038} (\bibinfo {year}
  {2004})}\BibitemShut {NoStop}%
\bibitem [{\citenamefont {Carberry}\ \emph {et~al.}(2004)\citenamefont
  {Carberry}, \citenamefont {Reid}, \citenamefont {Wang}, \citenamefont
  {Sevick}, \citenamefont {Searles},\ and\ \citenamefont
  {Evans}}]{Carberry2004}%
  \BibitemOpen
  \bibfield  {author} {\bibinfo {author} {\bibfnamefont {D.~M.}\ \bibnamefont
  {Carberry}}, \bibinfo {author} {\bibfnamefont {J.~C.}\ \bibnamefont {Reid}},
  \bibinfo {author} {\bibfnamefont {G.~M.}\ \bibnamefont {Wang}}, \bibinfo
  {author} {\bibfnamefont {E.~M.}\ \bibnamefont {Sevick}}, \bibinfo {author}
  {\bibfnamefont {D.~J.}\ \bibnamefont {Searles}},\ and\ \bibinfo {author}
  {\bibfnamefont {D.~J.}\ \bibnamefont {Evans}},\ }\bibfield  {title} {\bibinfo
  {title} {{Fluctuations and irreversibility: An experimental demonstration of
  a second-law-like theorem using a colloidal particle held in an optical
  trap}},\ }\href {https://doi.org/10.1103/PhysRevLett.92.140601} {\bibfield
  {journal} {\bibinfo  {journal} {Phys. Rev. Lett.}\ }\textbf {\bibinfo
  {volume} {92}},\ \bibinfo {pages} {140601} (\bibinfo {year}
  {2004})}\BibitemShut {NoStop}%
\bibitem [{\citenamefont {Alemany}\ \emph {et~al.}(2012)\citenamefont
  {Alemany}, \citenamefont {Mossa}, \citenamefont {Junier},\ and\ \citenamefont
  {Ritort}}]{Alemany2012}%
  \BibitemOpen
  \bibfield  {author} {\bibinfo {author} {\bibfnamefont {A.}~\bibnamefont
  {Alemany}}, \bibinfo {author} {\bibfnamefont {A.}~\bibnamefont {Mossa}},
  \bibinfo {author} {\bibfnamefont {I.}~\bibnamefont {Junier}},\ and\ \bibinfo
  {author} {\bibfnamefont {F.}~\bibnamefont {Ritort}},\ }\bibfield  {title}
  {\bibinfo {title} {{Experimental free-energy measurements of kinetic
  molecular states using fluctuation theorems}},\ }\href
  {http://dx.doi.org/10.1038/nphys2375} {\bibfield  {journal} {\bibinfo
  {journal} {Nat. Phys.}\ }\textbf {\bibinfo {volume} {8}},\ \bibinfo {pages}
  {688} (\bibinfo {year} {2012})}\BibitemShut {NoStop}%
\bibitem [{\citenamefont {Hoang}\ \emph {et~al.}(2018)\citenamefont {Hoang},
  \citenamefont {Pan}, \citenamefont {Ahn}, \citenamefont {Bang}, \citenamefont
  {Quan},\ and\ \citenamefont {Li}}]{Hoang2018a}%
  \BibitemOpen
  \bibfield  {author} {\bibinfo {author} {\bibfnamefont {T.~M.}\ \bibnamefont
  {Hoang}}, \bibinfo {author} {\bibfnamefont {R.}~\bibnamefont {Pan}}, \bibinfo
  {author} {\bibfnamefont {J.}~\bibnamefont {Ahn}}, \bibinfo {author}
  {\bibfnamefont {J.}~\bibnamefont {Bang}}, \bibinfo {author} {\bibfnamefont
  {H.~T.}\ \bibnamefont {Quan}},\ and\ \bibinfo {author} {\bibfnamefont
  {T.}~\bibnamefont {Li}},\ }\bibfield  {title} {\bibinfo {title} {Experimental
  test of the differential fluctuation theorem and a generalized {J}arzynski
  equality for arbitrary initial states},\ }\href
  {https://doi.org/10.1103/PhysRevLett.120.080602} {\bibfield  {journal}
  {\bibinfo  {journal} {Phys. Rev. Lett.}\ }\textbf {\bibinfo {volume} {120}},\
  \bibinfo {pages} {080602} (\bibinfo {year} {2018})}\BibitemShut {NoStop}%
\bibitem [{\citenamefont {Hofmann}\ \emph {et~al.}(2016)\citenamefont
  {Hofmann}, \citenamefont {Maisi}, \citenamefont {R\"ossler}, \citenamefont
  {Basset}, \citenamefont {Kr\"ahenmann}, \citenamefont {M\"arki},
  \citenamefont {Ihn}, \citenamefont {Ensslin}, \citenamefont {Reichl},\ and\
  \citenamefont {Wegscheider}}]{Hofmann2016}%
  \BibitemOpen
  \bibfield  {author} {\bibinfo {author} {\bibfnamefont {A.}~\bibnamefont
  {Hofmann}}, \bibinfo {author} {\bibfnamefont {V.~F.}\ \bibnamefont {Maisi}},
  \bibinfo {author} {\bibfnamefont {C.}~\bibnamefont {R\"ossler}}, \bibinfo
  {author} {\bibfnamefont {J.}~\bibnamefont {Basset}}, \bibinfo {author}
  {\bibfnamefont {T.}~\bibnamefont {Kr\"ahenmann}}, \bibinfo {author}
  {\bibfnamefont {P.}~\bibnamefont {M\"arki}}, \bibinfo {author} {\bibfnamefont
  {T.}~\bibnamefont {Ihn}}, \bibinfo {author} {\bibfnamefont {K.}~\bibnamefont
  {Ensslin}}, \bibinfo {author} {\bibfnamefont {C.}~\bibnamefont {Reichl}},\
  and\ \bibinfo {author} {\bibfnamefont {W.}~\bibnamefont {Wegscheider}},\
  }\bibfield  {title} {\bibinfo {title} {Equilibrium free energy measurement of
  a confined electron driven out of equilibrium},\ }\href
  {https://doi.org/10.1103/PhysRevB.93.035425} {\bibfield  {journal} {\bibinfo
  {journal} {Phys. Rev. B}\ }\textbf {\bibinfo {volume} {93}},\ \bibinfo
  {pages} {035425} (\bibinfo {year} {2016})}\BibitemShut {NoStop}%
\bibitem [{\citenamefont {Hofmann}\ \emph {et~al.}(2017)\citenamefont
  {Hofmann}, \citenamefont {Maisi}, \citenamefont {Basset}, \citenamefont
  {Reichl}, \citenamefont {Wegscheider}, \citenamefont {Ihn}, \citenamefont
  {Ensslin},\ and\ \citenamefont {Jarzynski}}]{Hofmann2017}%
  \BibitemOpen
  \bibfield  {author} {\bibinfo {author} {\bibfnamefont {A.}~\bibnamefont
  {Hofmann}}, \bibinfo {author} {\bibfnamefont {V.~F.}\ \bibnamefont {Maisi}},
  \bibinfo {author} {\bibfnamefont {J.}~\bibnamefont {Basset}}, \bibinfo
  {author} {\bibfnamefont {C.}~\bibnamefont {Reichl}}, \bibinfo {author}
  {\bibfnamefont {W.}~\bibnamefont {Wegscheider}}, \bibinfo {author}
  {\bibfnamefont {T.}~\bibnamefont {Ihn}}, \bibinfo {author} {\bibfnamefont
  {K.}~\bibnamefont {Ensslin}},\ and\ \bibinfo {author} {\bibfnamefont
  {C.}~\bibnamefont {Jarzynski}},\ }\bibfield  {title} {\bibinfo {title} {Heat
  dissipation and fluctuations in a driven quantum dot},\ }\href
  {https://doi.org/10.1002/pssb.201600546} {\bibfield  {journal} {\bibinfo
  {journal} {Phys. Status Solidi B}\ }\textbf {\bibinfo {volume} {254}},\
  \bibinfo {pages} {1600546} (\bibinfo {year} {2017})}\BibitemShut {NoStop}%
\bibitem [{\citenamefont {Liphardt}\ \emph {et~al.}(2001)\citenamefont
  {Liphardt}, \citenamefont {Onoa}, \citenamefont {Smith}, \citenamefont
  {Tinoco},\ and\ \citenamefont {Bustamante}}]{Liphardt2001}%
  \BibitemOpen
  \bibfield  {author} {\bibinfo {author} {\bibfnamefont {J.}~\bibnamefont
  {Liphardt}}, \bibinfo {author} {\bibfnamefont {B.}~\bibnamefont {Onoa}},
  \bibinfo {author} {\bibfnamefont {S.~B.}\ \bibnamefont {Smith}}, \bibinfo
  {author} {\bibfnamefont {I.}~\bibnamefont {Tinoco}},\ and\ \bibinfo {author}
  {\bibfnamefont {C.}~\bibnamefont {Bustamante}},\ }\bibfield  {title}
  {\bibinfo {title} {{Reversible unfolding of single RNA molecules by
  mechanical force}},\ }\href {https://doi.org/10.1126/science.1058498}
  {\bibfield  {journal} {\bibinfo  {journal} {Science}\ }\textbf {\bibinfo
  {volume} {292}},\ \bibinfo {pages} {733} (\bibinfo {year}
  {2001})}\BibitemShut {NoStop}%
\bibitem [{\citenamefont {Ritort}\ \emph {et~al.}(2002)\citenamefont {Ritort},
  \citenamefont {Bustamante},\ and\ \citenamefont {Tinoco}}]{Ritort2002}%
  \BibitemOpen
  \bibfield  {author} {\bibinfo {author} {\bibfnamefont {F.}~\bibnamefont
  {Ritort}}, \bibinfo {author} {\bibfnamefont {C.}~\bibnamefont {Bustamante}},\
  and\ \bibinfo {author} {\bibfnamefont {I.}~\bibnamefont {Tinoco}},\
  }\bibfield  {title} {\bibinfo {title} {{A two-state kinetic model for the
  unfolding of single molecules by mechanical force}},\ }\href
  {https://doi.org/10.1073/pnas.172525099} {\bibfield  {journal} {\bibinfo
  {journal} {Proc. Natl. Acad. Sci.}\ }\textbf {\bibinfo {volume} {99}},\
  \bibinfo {pages} {13544} (\bibinfo {year} {2002})}\BibitemShut {NoStop}%
\bibitem [{\citenamefont {Liphardt}(2002)}]{Liphardt2002}%
  \BibitemOpen
  \bibfield  {author} {\bibinfo {author} {\bibfnamefont {J.}~\bibnamefont
  {Liphardt}},\ }\bibfield  {title} {\bibinfo {title} {{Equilibrium Information
  from Nonequilibrium Measurements in an Experimental Test of Jarzynski's
  Equality}},\ }\href
  {http://www.sciencemag.org/cgi/doi/10.1126/science.1071152} {\bibfield
  {journal} {\bibinfo  {journal} {Science}\ }\textbf {\bibinfo {volume}
  {296}},\ \bibinfo {pages} {1832} (\bibinfo {year} {2002})}\BibitemShut
  {NoStop}%
\bibitem [{\citenamefont {Collin}\ \emph {et~al.}(2005)\citenamefont {Collin},
  \citenamefont {Ritort}, \citenamefont {Jarzynski}, \citenamefont {Smith},
  \citenamefont {Tinoco},\ and\ \citenamefont {Bustamante}}]{Collin2005}%
  \BibitemOpen
  \bibfield  {author} {\bibinfo {author} {\bibfnamefont {D.}~\bibnamefont
  {Collin}}, \bibinfo {author} {\bibfnamefont {F.}~\bibnamefont {Ritort}},
  \bibinfo {author} {\bibfnamefont {C.}~\bibnamefont {Jarzynski}}, \bibinfo
  {author} {\bibfnamefont {S.~B.}\ \bibnamefont {Smith}}, \bibinfo {author}
  {\bibfnamefont {I.}~\bibnamefont {Tinoco}},\ and\ \bibinfo {author}
  {\bibfnamefont {C.}~\bibnamefont {Bustamante}},\ }\bibfield  {title}
  {\bibinfo {title} {{Verification of the Crooks fluctuation theorem and
  recovery of RNA folding free energies}},\ }\href
  {https://doi.org/10.1038/nature04061} {\bibfield  {journal} {\bibinfo
  {journal} {Nature}\ }\textbf {\bibinfo {volume} {437}},\ \bibinfo {pages}
  {231} (\bibinfo {year} {2005})}\BibitemShut {NoStop}%
\bibitem [{\citenamefont {Manosas}\ and\ \citenamefont
  {Ritort}(2005)}]{Manosas2005}%
  \BibitemOpen
  \bibfield  {author} {\bibinfo {author} {\bibfnamefont {M.}~\bibnamefont
  {Manosas}}\ and\ \bibinfo {author} {\bibfnamefont {F.}~\bibnamefont
  {Ritort}},\ }\bibfield  {title} {\bibinfo {title} {{Thermodynamic and kinetic
  aspects of RNA pulling experiments}},\ }\href
  {http://dx.doi.org/10.1529/biophysj.104.045344} {\bibfield  {journal}
  {\bibinfo  {journal} {Biophys. J.}\ }\textbf {\bibinfo {volume} {88}},\
  \bibinfo {pages} {3224} (\bibinfo {year} {2005})}\BibitemShut {NoStop}%
\bibitem [{\citenamefont {Mossa}\ \emph {et~al.}(2009)\citenamefont {Mossa},
  \citenamefont {Manosas}, \citenamefont {Forns}, \citenamefont {Huguet},\ and\
  \citenamefont {Ritort}}]{ManosasMossa2009a}%
  \BibitemOpen
  \bibfield  {author} {\bibinfo {author} {\bibfnamefont {A.}~\bibnamefont
  {Mossa}}, \bibinfo {author} {\bibfnamefont {M.}~\bibnamefont {Manosas}},
  \bibinfo {author} {\bibfnamefont {N.}~\bibnamefont {Forns}}, \bibinfo
  {author} {\bibfnamefont {J.~M.}\ \bibnamefont {Huguet}},\ and\ \bibinfo
  {author} {\bibfnamefont {F.}~\bibnamefont {Ritort}},\ }\bibfield  {title}
  {\bibinfo {title} {{Dynamic force spectroscopy of DNA hairpins: I. Force
  kinetics and free energy landscapes}},\ }\href
  {https://doi.org/10.1088/1742-5468/2009/02/P02060} {\bibfield  {journal}
  {\bibinfo  {journal} {J. Stat. Mech. Theor. Exp.}\ }\textbf {\bibinfo
  {volume} {2009}},\ \bibinfo {pages} {P02060} (\bibinfo {year}
  {2009})}\BibitemShut {NoStop}%
\bibitem [{\citenamefont {Manosas}\ \emph {et~al.}(2009)\citenamefont
  {Manosas}, \citenamefont {Mossa}, \citenamefont {Forns}, \citenamefont
  {Huguet},\ and\ \citenamefont {Ritort}}]{ManosasMossa2009b}%
  \BibitemOpen
  \bibfield  {author} {\bibinfo {author} {\bibfnamefont {M.}~\bibnamefont
  {Manosas}}, \bibinfo {author} {\bibfnamefont {A.}~\bibnamefont {Mossa}},
  \bibinfo {author} {\bibfnamefont {N.}~\bibnamefont {Forns}}, \bibinfo
  {author} {\bibfnamefont {J.~M.}\ \bibnamefont {Huguet}},\ and\ \bibinfo
  {author} {\bibfnamefont {F.}~\bibnamefont {Ritort}},\ }\bibfield  {title}
  {\bibinfo {title} {{Dynamic force spectroscopy of DNA hairpins: II.
  Irreversibility and dissipation}},\ }\href
  {https://doi.org/10.1088/1742-5468/2009/02/P02061} {\bibfield  {journal}
  {\bibinfo  {journal} {J. Stat. Mech. Theor. Exp.}\ }\textbf {\bibinfo
  {volume} {2009}},\ \bibinfo {pages} {P02061} (\bibinfo {year}
  {2009})}\BibitemShut {NoStop}%
\bibitem [{\citenamefont {Dieterich}\ \emph {et~al.}(2015)\citenamefont
  {Dieterich}, \citenamefont {Camunas-Soler}, \citenamefont
  {Ribezzi-Crivellari}, \citenamefont {Seifert},\ and\ \citenamefont
  {Ritort}}]{Dieterich2015}%
  \BibitemOpen
  \bibfield  {author} {\bibinfo {author} {\bibfnamefont {E.}~\bibnamefont
  {Dieterich}}, \bibinfo {author} {\bibfnamefont {J.}~\bibnamefont
  {Camunas-Soler}}, \bibinfo {author} {\bibfnamefont {M.}~\bibnamefont
  {Ribezzi-Crivellari}}, \bibinfo {author} {\bibfnamefont {U.}~\bibnamefont
  {Seifert}},\ and\ \bibinfo {author} {\bibfnamefont {F.}~\bibnamefont
  {Ritort}},\ }\bibfield  {title} {\bibinfo {title} {{Single-molecule
  measurement of the effective temperature in non-equilibrium steady states}},\
  }\href {https://doi.org/10.1038/NPHYS3435} {\bibfield  {journal} {\bibinfo
  {journal} {Nat. Phys.}\ }\textbf {\bibinfo {volume} {11}},\ \bibinfo {pages}
  {971} (\bibinfo {year} {2015})}\BibitemShut {NoStop}%
\bibitem [{\citenamefont {Dieterich}\ \emph {et~al.}(2016)\citenamefont
  {Dieterich}, \citenamefont {Camunas-Soler}, \citenamefont
  {Ribezzi-Crivellari}, \citenamefont {Seifert},\ and\ \citenamefont
  {Ritort}}]{Dieterich2016}%
  \BibitemOpen
  \bibfield  {author} {\bibinfo {author} {\bibfnamefont {E.}~\bibnamefont
  {Dieterich}}, \bibinfo {author} {\bibfnamefont {J.}~\bibnamefont
  {Camunas-Soler}}, \bibinfo {author} {\bibfnamefont {M.}~\bibnamefont
  {Ribezzi-Crivellari}}, \bibinfo {author} {\bibfnamefont {U.}~\bibnamefont
  {Seifert}},\ and\ \bibinfo {author} {\bibfnamefont {F.}~\bibnamefont
  {Ritort}},\ }\bibfield  {title} {\bibinfo {title} {Control of force through
  feedback in small driven systems},\ }\href
  {https://doi.org/10.1103/PhysRevE.94.012107} {\bibfield  {journal} {\bibinfo
  {journal} {Phys. Rev. E}\ }\textbf {\bibinfo {volume} {94}},\ \bibinfo
  {pages} {012107} (\bibinfo {year} {2016})}\BibitemShut {NoStop}%
\bibitem [{\citenamefont {Ribezzi-Crivellari}\ and\ \citenamefont
  {Ritort}(2019{\natexlab{a}})}]{ribezzi:2019nat}%
  \BibitemOpen
  \bibfield  {author} {\bibinfo {author} {\bibfnamefont {M.}~\bibnamefont
  {Ribezzi-Crivellari}}\ and\ \bibinfo {author} {\bibfnamefont
  {F.}~\bibnamefont {Ritort}},\ }\bibfield  {title} {\bibinfo {title} {Large
  work extraction and the landauer limit in a continuous maxwell demon},\
  }\href {https://doi.org/10.1038/s41567-019-0481-0} {\bibfield  {journal}
  {\bibinfo  {journal} {Nat. Phys.}\ }\textbf {\bibinfo {volume} {15}},\
  \bibinfo {pages} {660} (\bibinfo {year} {2019}{\natexlab{a}})}\BibitemShut
  {NoStop}%
\bibitem [{\citenamefont {Ribezzi-Crivellari}\ and\ \citenamefont
  {Ritort}(2019{\natexlab{b}})}]{ribezzi:2019}%
  \BibitemOpen
  \bibfield  {author} {\bibinfo {author} {\bibfnamefont {M.}~\bibnamefont
  {Ribezzi-Crivellari}}\ and\ \bibinfo {author} {\bibfnamefont
  {F.}~\bibnamefont {Ritort}},\ }\bibfield  {title} {\bibinfo {title} {Work
  extraction, information-content and the landauer bound in the continuous
  maxwell demon},\ }\href {https://doi.org/10.1088/1742-5468/ab3340} {\bibfield
   {journal} {\bibinfo  {journal} {J. Stat. Mech. Theory Exp.}\ }\textbf
  {\bibinfo {volume} {2019}},\ \bibinfo {pages} {084013} (\bibinfo {year}
  {2019}{\natexlab{b}})}\BibitemShut {NoStop}%
\bibitem [{\citenamefont {Rico-Pasto}\ \emph {et~al.}(2021)\citenamefont
  {Rico-Pasto}, \citenamefont {Schmitt}, \citenamefont {Ribezzi-Crivellari},
  \citenamefont {Parrondo}, \citenamefont {Linke}, \citenamefont {Johansson},\
  and\ \citenamefont {Ritort}}]{rico:2020}%
  \BibitemOpen
  \bibfield  {author} {\bibinfo {author} {\bibfnamefont {M.}~\bibnamefont
  {Rico-Pasto}}, \bibinfo {author} {\bibfnamefont {R.~K.}\ \bibnamefont
  {Schmitt}}, \bibinfo {author} {\bibfnamefont {M.}~\bibnamefont
  {Ribezzi-Crivellari}}, \bibinfo {author} {\bibfnamefont {J.~M.~R.}\
  \bibnamefont {Parrondo}}, \bibinfo {author} {\bibfnamefont {H.}~\bibnamefont
  {Linke}}, \bibinfo {author} {\bibfnamefont {J.}~\bibnamefont {Johansson}},\
  and\ \bibinfo {author} {\bibfnamefont {F.}~\bibnamefont {Ritort}},\
  }\bibfield  {title} {\bibinfo {title} {Thermodynamic information and
  information-to-measurement conversion in {DNA} pulling experiments},\
  }\href@noop {} {\bibfield  {journal} {\bibinfo  {journal} {Phys. Rev. X}\
  }\textbf {\bibinfo {volume} {11}},\ \bibinfo {pages} {031052} (\bibinfo
  {year} {2021})}\BibitemShut {NoStop}%
\bibitem [{\citenamefont {Ritort}(2004)}]{Ritort2004}%
  \BibitemOpen
  \bibfield  {author} {\bibinfo {author} {\bibfnamefont {F.}~\bibnamefont
  {Ritort}},\ }\bibfield  {title} {\bibinfo {title} {Work and heat fluctuations
  in two-state systems: a trajectory thermodynamics formalism},\ }\href
  {http://stacks.iop.org/1742-5468/2004/i=10/a=P10016} {\bibfield  {journal}
  {\bibinfo  {journal} {J. Stat. Mech. Theor. Exp.}\ }\textbf {\bibinfo
  {volume} {2004}},\ \bibinfo {pages} {P10016} (\bibinfo {year}
  {2004})}\BibitemShut {NoStop}%
\bibitem [{\citenamefont {Chvosta}\ \emph {et~al.}(2007)\citenamefont
  {Chvosta}, \citenamefont {Reineker},\ and\ \citenamefont
  {Schulz}}]{Chvosta2007}%
  \BibitemOpen
  \bibfield  {author} {\bibinfo {author} {\bibfnamefont {P.}~\bibnamefont
  {Chvosta}}, \bibinfo {author} {\bibfnamefont {P.}~\bibnamefont {Reineker}},\
  and\ \bibinfo {author} {\bibfnamefont {M.}~\bibnamefont {Schulz}},\
  }\bibfield  {title} {\bibinfo {title} {Probability distribution of work done
  on a two-level system during a nonequilibrium isothermal process},\ }\href
  {https://link.aps.org/doi/10.1103/PhysRevE.75.041124} {\bibfield  {journal}
  {\bibinfo  {journal} {Phys. Rev. E}\ }\textbf {\bibinfo {volume} {75}},\
  \bibinfo {pages} {041124} (\bibinfo {year} {2007})}\BibitemShut {NoStop}%
\bibitem [{\citenamefont {\v{S}ubrt}\ and\ \citenamefont
  {Chvosta}(2007)}]{Subrt2007}%
  \BibitemOpen
  \bibfield  {author} {\bibinfo {author} {\bibfnamefont {E.}~\bibnamefont
  {\v{S}ubrt}}\ and\ \bibinfo {author} {\bibfnamefont {P.}~\bibnamefont
  {Chvosta}},\ }\bibfield  {title} {\bibinfo {title} {Exact analysis of work
  fluctuations in two-level systems},\ }\href
  {http://stacks.iop.org/1742-5468/2007/i=09/a=P09019} {\bibfield  {journal}
  {\bibinfo  {journal} {J. Stat. Mech. Theor. Exp.}\ }\textbf {\bibinfo
  {volume} {2007}},\ \bibinfo {pages} {P09019} (\bibinfo {year}
  {2007})}\BibitemShut {NoStop}%
\bibitem [{\citenamefont {Manosas}\ \emph {et~al.}(2007)\citenamefont
  {Manosas}, \citenamefont {Wen}, \citenamefont {Li}, \citenamefont {Smith},
  \citenamefont {Bustamante}, \citenamefont {Tinoco},\ and\ \citenamefont
  {Ritort}}]{Manosas2007b}%
  \BibitemOpen
  \bibfield  {author} {\bibinfo {author} {\bibfnamefont {M.}~\bibnamefont
  {Manosas}}, \bibinfo {author} {\bibfnamefont {J.-D.}\ \bibnamefont {Wen}},
  \bibinfo {author} {\bibfnamefont {P.~T.~X.}\ \bibnamefont {Li}}, \bibinfo
  {author} {\bibfnamefont {S.~B.}\ \bibnamefont {Smith}}, \bibinfo {author}
  {\bibfnamefont {C.}~\bibnamefont {Bustamante}}, \bibinfo {author}
  {\bibfnamefont {I.}~\bibnamefont {Tinoco}},\ and\ \bibinfo {author}
  {\bibfnamefont {F.}~\bibnamefont {Ritort}},\ }\bibfield  {title} {\bibinfo
  {title} {{Force unfolding kinetics of RNA using optical tweezers. II.
  Modeling experiments}},\ }\href {https://doi.org/10.1529/biophysj.106.094243}
  {\bibfield  {journal} {\bibinfo  {journal} {Biophys. J.}\ }\textbf {\bibinfo
  {volume} {92}},\ \bibinfo {pages} {3010} (\bibinfo {year}
  {2007})}\BibitemShut {NoStop}%
\bibitem [{\citenamefont {Alemany}(2014)}]{AlemanyThesis}%
  \BibitemOpen
  \bibfield  {author} {\bibinfo {author} {\bibfnamefont {A.}~\bibnamefont
  {Alemany}},\ }\emph {\bibinfo {title} {{Dynamic force spectroscopy and
  folding kinetics in molecular systems}}},\ \href
  {http://hdl.handle.net/2445/60025} {Ph.D. thesis},\ \bibinfo  {school}
  {Universitat de Barcelona} (\bibinfo {year} {2014})\BibitemShut {NoStop}%
\bibitem [{SI()}]{SI}%
  \BibitemOpen
  \href@noop {} {\bibinfo {title} {See supplementary information for
  details}}\BibitemShut {NoStop}%
\bibitem [{dis()}]{discrete}%
  \BibitemOpen
  \href@noop {} {\bibinfo {title} {We note that in the literature, discrete
  feedback is often denoting the single measurement case only}}\BibitemShut
  {NoStop}%
\bibitem [{\citenamefont {Alemany}\ and\ \citenamefont
  {Ritort}(2014)}]{Alemany2014}%
  \BibitemOpen
  \bibfield  {author} {\bibinfo {author} {\bibfnamefont {A.}~\bibnamefont
  {Alemany}}\ and\ \bibinfo {author} {\bibfnamefont {F.}~\bibnamefont
  {Ritort}},\ }\bibfield  {title} {\bibinfo {title} {Determination of the
  elastic properties of short ssdna molecules by mechanically folding and
  unfolding dna hairpins.},\ }\href@noop {} {\bibfield  {journal} {\bibinfo
  {journal} {Biopolymers}\ }\textbf {\bibinfo {volume} {101}},\ \bibinfo
  {pages} {1193} (\bibinfo {year} {2014})}\BibitemShut {NoStop}%
\end{thebibliography}%

\newpage
\widetext
\begin{center}
	\textbf{\large Supplemental information}
\end{center}
\setcounter{equation}{0}
\setcounter{figure}{0}
\setcounter{table}{0}
\setcounter{page}{1}
\makeatletter
\renewcommand{\theequation}{S\arabic{equation}}
\renewcommand{\thefigure}{S\arabic{figure}}

The Supplementary Information is structured as follows. In Sec.~I, we give detailed derivations of the work probability distribution in the absence of feedback. In Sec.~II, we provide background on the detailed fluctuation theorem used in the main text. In Sec.~III we provide details on the Monte Carlo simulations discussed in the main text.  Equation and Figure numbers not preceded by an `$S$' refer to the main text.

\section{I. Work probability distribution without feedback}
\label{sec:worknf}
We derive here an expression for the work distribution for the protocol with no feedback (FB), $P_\text{nf}(w)$. Two different, formally equivalent, approaches are taken. First we discuss the fomal derivation of the full analytical solution, which we show obeys Crooks fluctuation theorem. The solution also allows us to derive the analytical expressions for $P_\text{nf}(w)$ in the limit of small and large $\zeta$, presented in the main text. Thereafter we present an approach which is numerically convenient and also gives access to the individual cumulants of the work distribution. This latter approach also forms the basis for numerical evaluations of the work distribution in cases with FB, descussed below. 

The starting point for both approaches is the rate equation for the state probabilities, conveniently written in matrix form as
\begin{equation}
\frac{d}{d\tau}\left( \begin{array}{c} P_\text F(\tau) \\ P_\text U(\tau) \end{array} \right)=\zeta \left(\begin{array}{cc} -e^{\tau} & e^{-\tau}  \\ e^{\tau} & -e^{-\tau} \end{array}  \right)\left( \begin{array}{c} P_\text F(\tau) \\ P_\text U(\tau) \end{array} \right).
\label{MEc}
\end{equation}
From the normalization condition $P_\text F(\tau)+P_\text U(\tau)=1$  we arrive at Eq. (1) in the main text.

\subsection{Full work probability distribution}
From the solution of Eq. (\ref{MEc}) follows straightforwardly a number of useful partial results. Given that we are in the folded state F at time $\tau_0$, the probability that we jump (for the first time) to the unfolded state U at time $\tau_1>\tau_0$ is given by
\begin{equation}
P_\text{FU}(\tau_0,\tau_1)=\zeta e^{\tau_1-\zeta (e^{\tau_1}-e^{\tau_0})}.
\end{equation}
In the same way, given that we are in the state U at time $\tau_1$, the probability that we jump (for the first time) to state F at time $\tau_2>\tau_1$ is given by
\begin{equation}
P_\text{UF}(\tau_1,\tau_2)=\zeta e^{-\tau_2+\zeta (e^{-\tau_2}-e^{-\tau_1})}.
\end{equation}
Also, given that we are in the state U at time $\tau_1$, the probability that we do not jump back to F for any $\tau>\tau_1$  is 
\begin{equation}
P_\text{UU}(\tau_1)=e^{-\zeta e^{-\tau_1}}.
\end{equation}
\subsubsection{Probability for a given number of transitions}
From these partial results we can then derive the probabilities that the system undergoes in total an odd number of transitions, starting with probability unity in state F at (effectively) time $\tau \rightarrow -\infty$ and ending with probability unity in state U at time $\tau \rightarrow \infty$. The probability that there is only one transition can be written as
\begin{equation}
\label{eq:ont}
P_1=\int_{-\infty}^{\infty} d\tau_0 P_\text{FU}(-\infty,\tau_0)P_\text{UU}(\tau_0)=\zeta \int_{-\infty}^{\infty} d\tau_0 f(\tau_0)=2\zeta K_1(2\zeta),
\end{equation}
where  $f(\tau)=\exp[\tau-2\zeta \cosh(\tau)]$ and $K_1(z)$ is a modified Bessel function. In the same way, the probability for three transitions is
\begin{eqnarray}
P_3&=&\int_{-\infty}^{\infty} d\tau_0\int_{\tau_0}^{\infty} d\tau_1 \int_{\tau_1}^{\infty} d\tau_2 P_\text{FU}(-\infty,\tau_0)P_\text{UF}(\tau_0,\tau_1)P_\text{FU}(\tau_1,\tau_2)P_\text{UU}(\tau_2) \nonumber \\
&=& \zeta^3 \int_{-\infty}^{\infty} d\tau_0\int_{\tau_0}^{\infty} d\tau_1 \int_{\tau_1}^{\infty} dx_2 \frac{f(\tau_0)f(\tau_2)}{f(\tau_1)}.
\end{eqnarray}
The general result for $2n+1$ transitions can thus be written on the compact form
\begin{equation}
P_{2n+1}=\zeta^{2n+1}\int_{-\infty}^{\infty} d\tau_0\int_{\tau_0}^{\infty} d\tau_1 .... \int_{\tau_{2n-1}}^{\infty} d\tau_{2n} \frac{f(\tau_0)f(\tau_2)...f(\tau_{2n})}{f(\tau_1)f(\tau_3)... f(\tau_{2n-1})}.
\end{equation}
By construction, it holds that $\sum_{n=0}^{\infty} P_{2n+1}=1$ for any $\zeta$. 

\subsubsection{Work probabilities}
We know that for every trajectory, the total work performed is equal to the heat dissipated to the bath, a consequence of the symmetry of the protocol. Hence, based on the expression for $P_{2n+1}$ we can write down the full work probability distribution as
\begin{eqnarray}
&&P_\text{nf}(w)=\sum_{n=0}^{\infty} \int_{-\infty}^{\infty} d\tau_0 \int_{\tau_0}^{\infty} d\tau_1 ... \int_{\tau_{2n-1}}^{\infty} d\tau_{2n} P_\text{FU}(-\infty,\tau_0) \nonumber \\
&\times& P_\text{UF}(\tau_0,\tau_1)....P_\text{FU}(\tau_{n-1},\tau_{n})P_\text{UU}(\tau_{2n}) \delta\left(w-2\sum_{m=0}^{2n}(-1)^m\tau_m\right),
\end{eqnarray}
where the $\delta$-function imposes the work done given the transitions occured at $\tau_0,...\tau_{2n}$. Inserting the expressions above we can write the integral
\begin{eqnarray}
P_\text{nf}(w)&=& e^{w/2}\sum_{n=0}^{\infty} \zeta^{2n+1} \int_{-\infty}^{\infty} d\tau_0 \int_{\tau_0}^{\infty} d\tau_1 ... \int_{\tau_{2n-1}}^{\infty} d\tau_{2n} \frac{g(\tau_0)g(\tau_2)...g(\tau_{2n})}{g(\tau_1)g(\tau_3)... g(\tau_{2n-1})}\delta\left(w-2\sum_{m=0}^{2n}(-1)^m\tau_m\right),
\end{eqnarray}
where $g(\tau)=\exp[-2\zeta \cosh \tau]$. By introducing new variables $y_0=\tau_0$ and $y_p=\tau_p-\tau_{p-1}$ for $p\geq 1$ and then carrying out the integral over $y_0$ we have
\begin{eqnarray}
&&P_\text{nf}(w)= e^{w/2}\sum_{n=0}^{\infty} \frac{\zeta^{2n+1}}{2} \int_{0}^{\infty} dy_1 \int_{0}^{\infty} dy_2 ... \int_{0}^{\infty} dy_{2n} \nonumber \\
&\times&  \frac{g\left(\frac{w}{2}-\Sigma_{2n}\right)g\left(\frac{w}{2}+y_1+y_2-\Sigma_{2n}\right)...g\left(\frac{w}{2}+y_1+...+ y_{2n}-\Sigma_{2n}\right)}{g\left(\frac{w}{2}+y_1-\Sigma_{2n}\right).....g\left(\frac{w}{2}+y_1+y_2+... y_{2n-1}-\Sigma_{2n}\right)},
\label{wprobdist}
\end{eqnarray}
where $\Sigma_{2n}=y_2+y_4+....y_{2n}$. Note that it holds that $\int dw P_\text{nf}(w)=\sum_{n=0}^{\infty} P_{2n+1}=1$ for any $\zeta$. 

\subsubsection{Crooks Fluctuation Theorem}
In order to show that the full work distribution fulfills the Crooks fluctuation theorem, we note that the integrand in Eq. (\ref{wprobdist}), that is the expression on the second line, is invariant under the joint transformation $w\leftrightarrow -w$ and $y_1 \leftrightarrow y_{2n}$,  $y_2 \leftrightarrow y_{2n-1}$,...,$y_{n-1}   \leftrightarrow y_{n}$. As a consequence, each multiple integral in the sum is invariant under $w\leftrightarrow -w$ and hence $P_\text{nf}(w)e^{-w/2}=P_\text{nf}(-w)e^{w/2}$, or equivalently $P_\text{nf}(w)/P_\text{nf}(-w)=e^{w}$. Our system hence obeys the Crooks fluctuation relation. Note that without feedback, the \textit{backward} experiment is the same as the \textit{forward} experiment because of the time-reversal symmetry of the protocol.

\subsubsection{Analytical expressions for $P_\text{nf}(w|\zeta \gg 1)$ and $P_\text{nf}(w|\zeta \ll 1)$}
We have unfortunately not been able to evaluate the multiple integrals in Eq. (\ref{wprobdist}) analytically for arbitrary $\zeta$. However, for the limits of small and large $\zeta$ we can find analytical expressions. For $\zeta\ll 1$, only a single transition takes place. In this case we can identify the lowest order term, $n=0$, in the sum in Eq. (\ref{wprobdist}) as $P_\text{nf}(w|\zeta\ll 1)$ and find 
\begin{align}
\label{eq:fastdrive}
P_\text{nf}(w|\zeta\ll 1)&=\frac{\zeta}{2} e^{w/2}\frac{g(w/2)}{P_1}=\frac{1}{4K_1(2\zeta)} e^{w/2-2\zeta \cosh(w/2)},
\end{align}
recovering Eq.~(4) in the main text. Here the probability for observing a single jump, $P_1$, is given in Eq.~\eqref{eq:ont}. Strictly speaking, the last equation is only valid for values of $\zeta$ where $P_1=1$ and where the last expression becomes a Gumbel distribution given by Eq.~(6) in the main text. However, in contrast to the Gumbel distribution, Eq.~\eqref{eq:fastdrive} explicitly fulfills the Crooks fluctuation theorem. In the quasistatic limit, $\zeta \gg 1$, we can expand the exponent in Eq. (\ref{wprobdist}) to second order in $w$ giving effectively a Gaussian approximation for the work distribution, as $P_\text{nf}(w|\zeta\gg 1) \propto e^{-\alpha(w-\langle w \rangle)^2}$. Knowing the average value $\langle w \rangle_\text{nf}=\pi/(4\zeta)$, derived below, and the above derived form $P_\text{nf}(w) \propto e^{w/2}h(\omega)$ with $h(\omega)=h(-\omega)$, we directly get
\begin{equation}
\label{eq:slowdrive}
P_\text{nf}(w|\zeta\gg 1)=\frac{\sqrt{\zeta}}{\pi}\exp\left[-\frac{\zeta}{\pi}\left(w-\frac{\pi}{4\zeta}\right)^2\right],
\end{equation}
which is Eq. (3) in the main text.

\subsection{Effective rate equation for work distribution}
We use, as above, that for every trajectory, the total work $w$ performed is equal to the heat $q$ dissipated to the bath. Within the approach taken here, it is convenient to analyze directly the distribution for $q$. The starting point is 
the conditional probabilities $P_\text F(\tau,q)$ and $P_\text U(\tau,q)$ to find the system in state F/U at time $\tau$, given that the heat $q$ has been dissipated. From the $q$-resolved rate equation corresponding to Eq.~(\ref{MEc}), by Fourier transforming with respect to $q$ we get the effective rate equation
\begin{equation}
\frac{d}{d\tau}\left( \begin{array}{c} P_\text F(\tau,\xi) \\ P_\text U(\tau,\xi) \end{array} \right)=\zeta \left(\begin{array}{cc} -e^{\tau} &e^{-\tau} e^{i2\xi \tau}   \\ e^{\tau}e^{-i2\xi \tau}  & -e^{-\tau} \end{array}  \right)\left( \begin{array}{c} P_\text F(\tau,\xi) \\ P_\text U(\tau,\xi) \end{array} \right) .
\label{CGFq}
\end{equation}
where $\xi$ is the conjugated variable to $q$. The probability distribution $P(\tau,q)=P_\text F(\tau,q)+P_\text U(\tau,q)$ is given by
\begin{equation}
P(\tau,q)=\frac{1}{2\pi}\int_{-\infty}^{\infty}d\xi e^{-iq\xi}P(\tau,\xi), \hspace{0.3cm}  P(\tau,\xi)=P_\text F(\tau,\xi)+P_\text U(\tau,\xi).
\label{MGF}
\end{equation}
Starting in state F at $\tau \rightarrow -\infty$, we find the sought work distribution $P_\text{nf}(w)=P(\tau \rightarrow \infty,q=-w)$.

\subsubsection{Moment expansion, average work}
It follows by definition from Eq. (\ref{MGF}) that $P(\tau,\xi)$ is the moment generating function for the probability distribution $P(\tau,q)$. The components of the moment generating function can be expanded in $\xi$ as 
\begin{equation}
P_\text F(\tau,\xi)=P_\text F^{(0)}(\tau)+i\xi P_\text F^{(1)}(\tau)+...., \quad P_\text U(\tau,\xi)=P_\text U^{(0)}(\tau)+i\xi P_\text U^{(1)}(\tau)+...., 
\end{equation}
where we for shortness write $P_\text F^{(0)}(\tau)=P_\text F(\tau,0), P_\text F^{(1)}(\tau)=dP_\text F(\tau,\xi)/d(i\xi)|_{\xi=0}$ and similar for $P_\text U^{(n)}(\tau)$. By adding the two components we get for the moment generating function 
\begin{equation}
P(\tau,\xi)=P^{(0)}(\tau)+i\xi P^{(1)}(\tau)+....=1+i\xi \langle q(\tau)\rangle+....
\label{qexp}
\end{equation}
where we used the normalization $P_F^{(0)}(\tau)+P_U^{(0)}(\tau)\equiv P_F(\tau)+P_U(\tau)=1$ and $q(\tau)$ is the dissipated heat at time $\tau$. To find $ \langle q(\tau)\rangle_\text{nf} = P^{(1)}(\tau)$
we expand Eq. (\ref{CGFq}) to first order in $\xi$ giving
\begin{eqnarray}
\frac{d}{d\tau}\left( \begin{array}{c} P_\text F^{(1)}(\tau) \\ P_ \text U^{(1)}(\tau) \end{array} \right)&=&\zeta \left(\begin{array}{cc} -e^{\tau} &e^{-\tau}    \\ e^{\tau}  & -e^{-\tau} \end{array}  \right)\left( \begin{array}{c} P_\text F^{(1)}(\tau) \\ P_\text U^{(1)}(\tau) \end{array} \right) + \zeta \left(\begin{array}{cc} 0 &2\tau e^{-\tau}  \\ -2\tau e^{\tau}  & 0 \end{array}  \right)\left( \begin{array}{c} P_\text F^{(0)}(\tau) \\ P_\text U^{(0)}(\tau) \end{array} \right).
 \label{firstorder}
\end{eqnarray}
The zeroth order solutions, given from Eq.~(\ref{MEc}), are
\begin{eqnarray}
P_\text F^{(0)}(\tau)=\zeta \int_{-\infty}^{\tau} ~ds~ e^{-s+2\zeta[\sinh(s)-\sinh(\tau)]},  \quad P_\text U^{(0)}(\tau)=\zeta \int_{-\infty}^{\tau} ~ds~ e^{s+2\zeta[\sinh(s)-\sinh(\tau)]},
\end{eqnarray}
Adding the two equations for the components in (\ref{firstorder}) we thus get
\begin{eqnarray}
\frac{dP^{(1)}(\tau)}{d\tau}=\frac{d\langle q(\tau) \rangle_\text{nf}}{d\tau}=-2\tau \zeta \left[e^{\tau} P_\text F^{(0)}(\tau)-e^{-\tau}P_\text U^{(0)}(\tau)\right]=4\tau \zeta^2\int_{-\infty}^{\tau} ~ds~\sinh(s-\tau) e^{2\zeta[\sinh(s)-\sinh(\tau)]}.
\label{qder}
\end{eqnarray}
We thus directly get the work perfomed $\langle w \rangle_\text{nf}$ from $-\langle q(\infty)\rangle_\text{nf}$, or, by integrating up (\ref{qder}) as
\begin{eqnarray}
\label{eq:avwork}
\langle w \rangle_\text{nf}=\int_{-\infty}^{\infty} ~d\tau ~4\tau\zeta^2\int_{-\infty}^\tau ~ds~\sinh(s-\tau) e^{2\zeta[\sinh(s)-\sinh(\tau)]}=\frac{\pi^2}{4}\zeta \left[J_{0}(2\zeta)J_1(2\zeta)+Y_0(2\zeta)Y_1(2\zeta)\right].
\end{eqnarray}
were the last equality follows after some manipulations. This is Eq.~(5) in the main text. For small $\zeta \ll 1$, we get to leading order
\begin{eqnarray}
\langle w \rangle_\text{nf}=-\frac{1}{2}\left(\ln \zeta+\gamma_E\right)
\end{eqnarray}
in line with the result for continuous feedback, Eq.~(7) in the main text, as expected. 

\section{II. Detailed fluctuation theorem}
\label{sec:detfluc}
	This section is based on Ref.~\cite{potts:2018} and provides background on the detailed fluctuation theorem used in the main text [cf.~Eq.~(10) in the main text]. To this end, we label a given trajectory by $x$. A trajectory is specified by an instantaneous state U or F for each moment in time. In our limit of large initial (and final) level splitting $\Delta=\kappa\mathcal{T}\gg k_BT$, each trajectory starts in F and ends in U. We now consider protocols, where the energy levels are driven by some speed $\zeta$ (corresponding to $\kappa$) up to (dimensionless) time $\tau_s^*$, after which the speed is changed to $\zeta'$ (corresponding to $\kappa'$). Such a protocol will be denoted by $\lambda_s$. The level splitting at the moment of the velocity change is given by $\Delta_s=E_F(\tau_s^*)-E_U(\tau_s^*)=2k_BT\tau_s^*$. We also consider the time-reversed protocol ${\lambda}_s^\dagger$, where the system starts in state U with $E_F-E_U=\Delta$ (i.e., in thermal equilibrium) and is driven with the speed $\zeta'$ until the level spacing is equal to $E_F-E_U=\Delta_s$, before the speed is changed to $\zeta$. The protocols $\lambda_s$ will be applied in the \textit{forward experiment}, the protocols $\lambda_s^\dagger$ in the \textit{backward experiment}. For fixed protocols, our stochastic system obeys the well established detailed fluctuation theorem
	\begin{equation}
	\label{eq:detail0}
	\frac{P({x}^\dagger|\lambda_s^\dagger)}{P(x|\lambda_s)}=e^{-w(x)}\hspace{1cm}\Rightarrow\hspace{1cm}\frac{P(-w|\lambda_s^\dagger)}{P(w|\lambda_s)}=e^{-w},
	\end{equation}
	where $w(x)$ denotes the (dimensionless) work which is uniquely determined by the trajectory and daggered quantities are related to undaggered quantities by time-reversal. Here we made use of the fact that the free energy of the initial states are the same for the forward and the backward experiment. The fluctuation theorem for the work is simply obtained by summing over all trajectories which result in the same value of work. We stress that Eq.~\eqref{eq:detail0} holds no matter how it is decided that the protocol $\lambda_s$ is applied.
	
	We now consider the case of feedback, where $s$ corresponds to a measurement outcome. In our case it denotes the first measurement that gives the result U in the forward experiment. In this case, the joint probability distribution for $s$ and $x$ can be written as \cite{Sagawa2012}
	\begin{equation}
	\label{eq:jointproba}
	P(x,s)=P(x|\lambda_s)p_{s|x},
	\end{equation} 
	where $p_{s|x}$ denotes the probability of measuring $s$ for a fixed trajectory $x$. We further introduce the conditional probability 
	\begin{equation}
	\label{eq:condproba}
	P(x|s)=P(x,s)/p_s,\hspace{1cm}p_s=\int dxP(x|\lambda_s)p_{s|x},
	\end{equation}
	where $p_s$ denotes the probability of measuring $s$. We can now rewrite Eq.~\eqref{eq:detail0} as
	\begin{equation}
	\label{eq:detail1}
	\frac{{P}^\dagger({x}^\dagger|s)}{P(x|s)}\frac{{p}^\dagger_s}{p_s}=e^{-w(x)},
	\end{equation}
	where we introduced
	\begin{equation}
	\label{eq:backwardproba}
	{P}^\dagger(x^\dagger|s)=P(x^\dagger|\lambda_s^\dagger)\frac{p_{s|x}}{{p}^\dagger_s},\hspace{1cm}{p}^\dagger_s=\int d{x}P(x^\dagger|\lambda_s^\dagger)p_{s|x}.
	\end{equation}
	The backward conditional probability distribution can be understood as the conditional probability of the system to take trajectory $x^\dagger$, given that we apply protocol $\lambda_s^\dagger$ and post-select on measurement outcomes $s$ that comply with the applied protocol. Here we implicitly assume that the measurement outcome $s$ is equally probable on a forward experiment with trajectory $x$ and on a backward experiment with trajectory $x^\dagger$. Summing up all trajectories which result in the same work value and introducing $\Phi_s=\ln(p_s/{p}^\dagger_s)$ we obtain Eq.~(10) in the main text.
	
	We note that while $\sum_sp_s=1$ by construction, the backward probabilities $\sum_s{p}^\dagger_s=\gamma$ sum to the efficacy parameter. This is because in a forward experiment with feedback, the protocol depends on past measurement outcomes which depend on past system states. In a backward experiment, the protocol is fixed, i.e., no feedback is performed, and there is a finite probability that the measurement outcomes do not agree with the applied protocol.
	
\section{III. Monte Carlo simulations}
\label{sec:MC}

For the Monte Carlo simualtion, the unzipping of a short DNA hairpin ($20$ base pair (bp) stem plus a tetra loop) tethered between two polystyrene beads, one held with a micro pipette, the other trapped via 29 bp DNA handles with optical tweezers, is modeled as a Markov chain. The distance between the center of the optical trap and the micro pipette is the control parameter $\mathcal{L}$. Transitions between the natural folded and the unfolded state are defined through the attempt rate, the barrier height $B(\mathcal{L})$ and the free energy $\Delta F(\mathcal{L})$. In principle $B(\mathcal{L})$ and $\Delta F(\mathcal{L})$ are functions of the number of open base pairs and contain contributions of the handles, the linker-molecules and the bead of a typical optical tweezers setup \cite{Manosas2005,Manosas2007b,Huguet2010,Alemany2014}. However, short DNA hairpins unfold in a cooperative way \cite{AlemanyThesis} and can thus be simulated considering only transitions between the completely folded and the completely unfolded state, in analogy to the simple model considered above.
For each molecule 20k-100k trajectories (force, position, time) are simulated with time steps $10^{-4}$\,s. After subtraction of an equilibrium trajectory, work contributions are calculated as $W=\int_{l_0}^{l_1} d\mathcal{L} f(\mathcal{L})$ with $f(\mathcal{L})$ denoting the force acting on the molecule and $l_0~(l_1)$ the initial (final) control parameter. Using the transition statistics, $\zeta$ can be extracted. For more information, we refer to reader to the supplemental material of Ref~\cite{rico:2020}.

\end{document}